\documentclass[11pt,oneside,letter]{article}

\usepackage{amsthm,amsmath,amssymb,amsfonts,epsfig,mathrsfs,mathtools}
\usepackage{epstopdf}

\usepackage{graphicx}
\usepackage[font=sl,labelfont=bf]{caption}
\usepackage{subcaption}

\usepackage{slashbox}

\usepackage{cancel}

\usepackage{enumerate}

\usepackage[usenames,dvipsnames]{color}

\definecolor{labelkey}{rgb}{0,0,1}

\usepackage[colorlinks=true, pdfstartview=FitV, linkcolor=blue,
            citecolor=blue, urlcolor=blue]{hyperref}
\usepackage[usenames]{color}
\definecolor{Red}{rgb}{0.7,0,0.1}
\definecolor{Green}{rgb}{0,0.7,0}

\usepackage[numbers]{natbib}

\usepackage{url}
\makeatletter\def\url@leostyle{%
 \@ifundefined{selectfont}{\def\UrlFont{\sf}}{\def\UrlFont{\scriptsize\ttfamily}}} \makeatother

\usepackage{accents, comment}

\usepackage{bbm}

\usepackage{mathrsfs}

\newtheorem{theorem}{Theorem}
\newtheorem{proposition}[theorem]{Proposition}
\newtheorem{lemma}[theorem]{Lemma}
\newtheorem{corollary}[theorem]{Corollary}

\theoremstyle{definition}
\newtheorem{definition}[theorem]{Definition}
\newtheorem{example}[theorem]{Example}

\theoremstyle{remark}
\newtheorem{remark}[theorem]{Remark}

\numberwithin{equation}{section}
\numberwithin{theorem}{section}

\renewcommand{\mid}{\;|\;}              
\newcommand{\abs}[1]{\left\vert#1\right\vert}   
\renewcommand{\d}{\operatorname{d}\!}   

\newcommand{\overbar}[1]{\mkern 1.5mu\overline{\mkern-1.5mu#1\mkern-1.5mu}\mkern 1.5mu}

\newcommand{\stkout}[1]{\ifmmode\text{\sout{\ensuremath{#1}}}\else\sout{#1}\fi}

\usepackage{epstopdf}
\usepackage{tablefootnote}
\usepackage{array}
\usepackage{arydshln}
\usepackage{multirow}
\usepackage{cases}
\usepackage[normalem]{ulem}
\usepackage{kbordermatrix}
\allowdisplaybreaks

\DeclareSymbolFont{largesymbolsA}{U}{txexa}{m}{n}
\DeclareMathSymbol{\varprod}{\mathop}{largesymbolsA}{16}

\DeclareFontFamily{U}{mathx}{\hyphenchar\font45}
\DeclareFontShape{U}{mathx}{m}{n}{
      <5> <6> <7> <8> <9> <10>
      <10.95> <12> <14.4> <17.28> <20.74> <24.88>
      mathx10
      }{}
\DeclareSymbolFont{mathx}{U}{mathx}{m}{n}
\DeclareMathSymbol{\bigtimes}{1}{mathx}{"91}

\usepackage{epstopdf}
\usepackage{subcaption}     
\usepackage{tablefootnote}
\usepackage{array}
\usepackage{arydshln}
\usepackage{multirow}
\usepackage[titletoc]{appendix}

\usepackage{enumitem}
\usepackage{kbordermatrix}
\usepackage{comment}

\makeatletter
\newcommand*\rel@kern[1]{\kern#1\dimexpr\macc@kerna}
\newcommand*\widebar[1]{%
  \begingroup
  \def\mathaccent##1##2{%
    \rel@kern{0.8}%
    \overline{\rel@kern{-0.8}\macc@nucleus\rel@kern{0.2}}%
    \rel@kern{-0.2}%
  }%
  \macc@depth\@ne
  \let\math@bgroup\@empty \let\math@egroup\macc@set@skewchar
  \mathsurround\z@ \frozen@everymath{\mathgroup\macc@group\relax}%
  \macc@set@skewchar\relax
  \let\mathaccentV\macc@nested@a
  \macc@nested@a\relax111{#1}%
  \endgroup
}

\makeatletter
\def\namedlabel#1#2{\begingroup
    #2%
    \def\@currentlabel{#2}%
    \phantomsection\label{#1}\endgroup
}
\makeatother

\makeatletter
\newcommand{\labeltext}[2]{%
  \@bsphack
  \csname phantomsection\endcsname 
  \def\@currentlabel{#1}{\label{#2}}%
  \@esphack
}
\makeatother

\begin{document}
\title{\textit{Systemic Risk and the Dependence Structures}}
\author{Yu-Sin Chang\footnote{Postal address: Department of Mathematics, Milwaukee School of Engineering\newline
\hspace*{1.8em}1025 North Broadway, Milwaukee, WI 53202, USA\newline
\hspace*{1.8em}Email: \url{changy@msoe.edu}}}
\date{\today}
\maketitle

\begin{abstract}
We propose a dynamic model of dependence structure between financial institutions within a financial system and we construct measures for dependence and financial instability. Employing Markov structures of joint credit migrations, our model allows for contagious simultaneous jumps in credit ratings and provides flexibility in modeling dependence structures. Another key aspect is that the proposed measures consider the interdependence and reflect the changing economic landscape as financial institutions evolve over time. In the final part, we give several examples, where we study various dependence structures and investigate their systemic instability measures. In particular, we show that subject to the same pool of Markov chains, the simulated Markov structures with distinct dependence structures generate different sequences of systemic instability.\\





\end{abstract}


\tableofcontents

\section{Introduction}\label{sec:intro}


The most recent global financial crisis has highlighted interconnectedness in the financial system as a crucial source of systemic risk. In the 1990s, the concept of modeling systemic risk focused on the \textit{too-big-to-fail} issue, whereas the recent financial crisis addresses the \textit{too-connected-to-fail} problem. Since financial institutions are linked by financial activities, the default of one systemically important financial institution may jeopardize the stability of other financial counterparties. This may lead to successive rounds of failures, where the default of one financial institution escalates the financial distress of other engaged institutions. Under adverse circumstances, the distress may propagate through the whole financial system and cause the collapse of the financial system. On the other hand, a default of a ``bad player'' in a financial system may lead to improved health of the surviving part of the system. Such a phenomenon can be justifiably termed as ``systemic benefit''. 

We believe this financial phenomenon is highly related to the dependence structure within a financial system, namely, the interdependence between financial institutions. It is well recognized that understanding how the dependence structure will influence the financial stability of a financial system remains one of the key challenges faced by the market participants today.

The literature regarding various aspects of modeling systemic risk, in particular the network models, has been rapidly growing in the recent years. We refer the reader to some recent studies, e.g., \cite{am12sr, bc15interbank, ccvw14sr, eisenberg2001systemic, fouque2013handbook, glasserman2015likely, summer2013network} and the references therein. 

The main objective of this paper is to develop a novel methodology to study the dependence structure between financial institutions and the financial stability of the system, in a dynamic framework, and to test it via a numerical study. In particular, we propose new measures for computing the levels of dependence and financial instability of a financial economy. These measures account for credit migrations of the financial institutions and the stochastic dependence between these migrations, which is modeled in terms of Markov structures.

The concept of \textit{Markov structures}\footnote{In the previous works, a Markov structure was called a Markov coupla.} for multivariate Markov processes was originated in Bielecki et al. \cite{bvv2008jcr}. A Markov structure for a collection of Markov chains is a multivariate Markov chain whose components are Markov chains with identical probability laws to the prescribed Markov chains. In \cite{bjvv08dependence}, the authors have successfully applied the Markov structures theory to financial markets, in particular, to the basket-type products in credit risk. Among others, Cr\'{e}pey et al. \cite{cjz09cds} study the counterparty risk on a payer CDS based on a Markov structures approach; Bielecki et al. \cite{bcf18central} use a Markov structures model to investigate the dependence structure between the Central Clearing Parties. Whereas, these works confine their numerical studies to the strong Markov structures (cf. Definitions \ref{def:strongMarkovStruc} and Definition \ref{def:weakMarkovStruc} for weak Markov structures). Typically, for a strong Markov structure there is no contagious impact between the components while the Markovian components changing their states. In a nutshell, our proposed methodology is not restricted to the strong Markov structures. Another important aspect of the numerical study is to present the possibilities and accommodate the needs in modeling from the practical viewpoints.

Toward this end, we consider a financial economy consisting of finite numbers of financial institutions, where financial institutions are assigned credit migrating processes to reflect their financial stability. We assume that these individual credit migration processes are Markov chains. Next, we introduce a joint credit migration process whose components have the same probability laws as these individual credit migration processes, respectively. This joint credit migration process contains full information about the stochastic dependence between individual migrations. We propose to model this joint migration process as a Markov structure. Since the infinitesimal generator of a Markov structure encodes the dependence structure between these migrations, each Markov structure will characterize the evolution of dependence structure between the individual institutions. 

Furthermore, in an interconnected financial system, the migration of financial stability for one financial institution has direct or indirect impacts on the other financial institutions. Generally speaking, the phenomenon in which a shock from one financial institution is transmitted to another financial institution is called \textit{contagion}\footnote{In literature, there are serious debates on the definitions of \textit{contagion} and \textit{spillovers}. These definitions are model-dependent. Here, instead of giving precise definitions, we want to introduce the concept of the financial chain reactions.}. In our framework, this transmission mechanism can be introduced by imposing conditions on the generator of the Markov structure. That is, we consider weak-only Markov structures to model the joint credit migration process such that the credit migrations of some financial institutions are allowed to affect other credit migrations. 

Now, we shift our attention to measuring systemic risk and systemic stability. In literature, there are various perspectives on systemic measures, or what general properties a systemic measure needs to preserve. For instance, \cite{appr12msr,bff2015smr,bisias2012sr,chen2012systemic,frw2016sr,giglio10cds,hoffmann2016risk,se17consistent,segoviano09bsm}, and the reference therein.

Here we propose three measures to compute systemic risk, stochastic dependence between the financial institutions within a financial system, as well as systemic instability of this system. Let us start with the systemic risk measure. Suppose that a multivariate Markov chain $X$ taking finite values in $E$ is endowed with dependence structure $\mathcal{D}$ between its components; to emphasize this, we will use the notation $X^{\mathcal{D}}=(X^{\mathcal{D},1}\ldots,X^{\mathcal{D},m})$. Let $f$ be a real valued integrable function. We are interested in the conditional probabilities of of the form
\begin{equation}\label{eq:sdm-1}
\mathbb{P} \left( f\left( X^{\mathcal{D}}_T \right)\in B \mid X^{\mathcal{D}}_{t} =x\right) ,\quad x\in E,\ 0\leq t\leq T <\infty,
\end{equation}
where $B$ is a Borel set. We would like to compute the conditional probability of a fixed number of financial institutions that will be in certain credit ratings at some future time. 
This concept of measuring systemic risk is common in the financial literature. For instance, proposing the joint default probability for a systemic event. However, we argue that, using only the probability of systemic default events without considering the dependence structure between financial institutions is not thorough. 
It may happen that a financial system endowed with distinct dependence structures produces the same level of the joint default probability at some time $t$. We believe that the dependence structure is at the core of studying systemic risk and should be taken into account.

To account for the impact of dependence structure, we propose a systemic dependence measure of the form,
\begin{equation}\label{eq:sdm-2}
\mathbb{P} \left( f\left( X^{\mathcal{D}}_T\right)\in B \mid X^{\mathcal{D}}_{t}=x \right) -\mathbb{P} \left( f\left( X^{\mathcal{I}}_T\right)\in B \mid X^{\mathcal{I}}_{t}=x \right),\quad x\in E,\ 0\leq t\leq T<\infty,
\end{equation} 
where $\mathcal{I}$ stands for the independence structure. Namely, we normalize the systemic risk measure through the independence structure (cf. Definition \ref{def:sdm}). 
Under the Markov structures framework, since the Markov structures $X^{\mathcal{D}}$ and $X^{\mathcal{I}}$ are subject to the same pool of Markov chains, the normalization of \eqref{eq:sdm-2} has solid grounds. 

Additionally, we extend the systemic dependence measure to systemic instability measure to highlight the distance between distributions of $X_{t}^{\mathcal{D}}$ and $X_{t}^{\mathcal{I}}$ (cf. Definition \ref{def:sim}). Not only does the systemic instability measure include the ingredients of joint default probability and dependence, but it also allows us to track systemic instability for a fixed monitor window on a regular basis. 



The paper is organized as follows. In Section \ref{sec:preliminary} we introduce some notations and briefly recall related results to lay the foundations for our work. In Section \ref{sec:markovCon} we set forth the concept of Markovian consistency, which concerns the Markov property of components of a multivariate Markov chain in different filtrations. The Markovian consistency theory plays a crucial role in the Markov structures theory. Next, we introduce Markov structures in Section \ref{sec:markovStruc}, and provide a continuous time algorithm to solve for weak-only Markov structures. Section \ref{sec:depStru} is dedicated to a dynamic framework of modeling stochastic dependence between financial institutions. We explain the concept of dependence and the meaning of contagion in the context of a Markov structures model. Besides, a discrete-time algorithm for constructing weak-only Markov structures is provided. Right after the algorithm, a short discussion about the calibration of dependence structure is presented. In Section \ref{sec:sr}, we construct systemic risk measure, systemic dependence measure, and systemic instability measure as a gateway to the abstract concepts of dependence and financial stability. We explain financial meanings and examine the mathematical properties for the measures: invariant with respect to the permutation of the financial institutions and law-invariant. Finally, comprehensive numerical examples are given in Section \ref{sec:numerical}, where we provide various dependence structures and describe how to adopt this framework to monitor financial stability. 
For convenience, we defer to Appendix \ref{apd:results} some technical results related to this paper. 

\section{Preliminaries}\label{sec:preliminary}

Let $(\Omega ,\mathcal{F},\mathbb{P})$ be a complete underlying probability space. We consider a finite time horizon. Throughout this paper, for any process $Y$ defined on $(\Omega ,\mathcal{F},\mathbb{P})$, we denote by $\mathbb{F}^{Y}=(\mathcal{F}_{t}^{Y},t\geq 0)$ the natural filtration of $Y$, $\mathcal{F}_{t}^{Y}=\sigma\lbrace Y_{u},\ 0\leq u\leq t\rbrace$, and denote by $\mu^{Y}$ the initial distribution of $Y$. We fix a positive integer $m$, 
and consider a collection of finite sets $E_{i},\ i=1,2,\ldots ,m$. Next, let $E = E_{1}\times E_{2}\times \cdots \times E_{m}$ be the Cartesian product. The elements of $E$ are denoted by $z=\left( z^{1},\ldots ,z^{m}\right)$. 

We consider a continuous time multivariate Markov chain $X=\left( X^{1},\ldots ,X^{m}\right)$ defined on $(\Omega ,\mathcal{F},\mathbb{P})$ taking values in $E$. Thus, it holds that
\begin{equation}\label{eq:Markov-1}
\mathbb{P}\left( X_{t+u}\in\Gamma \mid \mathcal{F}_{t}^{X}\right) = \mathbb{P}\left( X_{t+u}\in\Gamma \mid X_{t}\right),\quad t,u\geq 0,\ \Gamma \subset E.
\end{equation}
Equivalently, the Markov property \eqref{eq:Markov-1} can be written as
\begin{equation}\label{eq:Markov-2}
\mathbb{P}\left( X_{t_{n+1}} =x_{n+1} \mid X_{t_{n}}=x_{n},\ X_{t_{n-1}}=x_{n-1},\ldots ,\ X_{0}=x_{0}\right) = \mathbb{P}\left( X_{t_{n+1}} =x_{n+1} \mid X_{t_{n}} =x_{n}\right),
\end{equation}
for every $n\in\mathbb{N}$, $0\leq t_1\leq t_2\leq \ldots \leq t_n\leq t_{n+1}$ and $x_{j}\in E$, $j=0,1,\ldots ,n+1$, and whenever the conditional probabilities on the both sides are well-defined.

The transition probability function of $X$ is denoted by $\mathbf{P}_{t,s}=\left[ \mathbf{P}_{t,s}^{x,y}\right]_{x,y\in E}, 0\leq t\leq s< \infty$, namely, $\mathbf{P}_{t,s}^{x,y}=\mathbb{P}(X_{s}=y\mid X_{t}=x)$. We assume that for every $t\geq 0$ the following limit exists outside of a set of Lebesgue zero,
\begin{equation}\label{def:gen}
\Lambda_{t} :=\lim_{h\downarrow 0}\frac{\mathbf{P}_{t,t+h}-\mathbf{I}_{\abs{E}\times\abs{E}}}{h},
\end{equation}
where $\mathbf{I}_{\abs{E}\times\abs{E}}$ is the identity matrix of dimension $\abs{E}$-by-$\abs{E}$, with $\abs{E}$ the cardinality of set $E$. On the set of Lebesgue zero we take $\Lambda_{t}=0$. The matrix function $\Lambda_{t}:= \left[ \lambda_{t}^{x,y}\right],\ 0\leq t<\infty$, defined in \eqref{def:gen} is called the infinitesimal generator function of $X$. Followed by \eqref{def:gen}, for all $x, y\in E$, entry $\lambda_{t}^{x,y}\geq 0$ for $x\neq y$, $\lambda_{t}^{x,x}\leq 0$, and
\begin{equation*}
\sum_{y\in E} \lambda_{t}^{x,y} =0 .
\end{equation*}
We will also postulate measurability and some mild integrable properties about $\Lambda_{t},\  0\leq t <\infty$. Therefore, the Kolmogorov forward equations are satisfied, in particular, in matrix notation,
\begin{equation*}
\frac{\partial}{\partial s} \mathbf{P}_{t,s} = \mathbf{P}_{t,s} \Lambda_{s} ,\quad \mathbf{P}_{t,t}=\mathbf{I}_{\abs{E}\times\abs{E}}.
\end{equation*}

\subsection{Markovian consistency}\label{sec:markovCon}
From here on, we sharpen our focus to two definitions of Markovian consistency in the context of the multivariate Markov chain $X$. Moreover, we collect some relevant concepts and results from \cite{bjn13intricacies, ysc17phd} to lay the basis for our work.
\begin{definition}[Strong Markovian consistency, {\cite[Definition 1.2]{bjn13intricacies}}]\label{def:smc}
A multivariate Markov chain $X$ is \textit{strongly Markovian consistent} relative to the component $X^{i}$ if
\begin{equation}\label{eq:smc}
\mathbb{P}\left( X_{s}^{i} \in \Gamma^{i} \mid \mathcal{F}_{t}^{X}\right) = \mathbb{P}\left( X_{s}^{i} \in \Gamma^{i} \mid X_{t}^{i}\right),
\end{equation}
for every $\Gamma^{i}\subset E_{i}$, $0\leq t\leq s<\infty$. If \eqref{eq:smc} holds for every $i\in\lbrace 1,2,\ldots ,m\rbrace$, then we say that $X$ satisfies the strong Markovian consistency property, or that $X$ is strongly Markovian consistent.
\end{definition}

Next concept is weaker than strong Markovian consistency.

\begin{definition}[Weak Markovian consistency, {\cite[Definition 1.1]{bjn13intricacies}}]\label{def:wmc}
A multivariate Markov chain $X$ is \textit{weakly Markovian consistent} relative to the component $X^{i}$ if
\begin{equation}\label{eq:wmc}
\mathbb{P}\left( X_{s}^{i} \in \Gamma^{i} \mid \mathcal{F}_{t}^{X^{i}}\right) = \mathbb{P}\left( X_{s}^{i} \in \Gamma^{i} \mid X_{t}^{i}\right),
\end{equation}
for every $\Gamma^{i}\subset E_{i}$, $0\leq t\leq s<\infty$.
If \eqref{eq:wmc} holds for every $i\in \lbrace 1,2,\ldots ,m\rbrace$, then we say that $X$ satisfies the weak Markovian consistency property, or that $X$ is weakly Markovian consistent.
\end{definition}

If a multivariate Markov chain is weakly Markovian consistent, but not strongly Markovian consistent, we say that this multivariate Markov chain satisfies the \textit{weak-only Markovian consistency} property.

\begin{definition}\label{def:Theta-Phi}
For any $i=1,2,\ldots ,m,$ and $t\geq 0$ the operator $\Theta_{t}^{i}$ acting on any function $g:E\rightarrow\mathbb{R}$ is defined as
\begin{equation*}\label{eq:ThetaOperator}
\Theta_{t}^{i}g \left( x^{i}\right) =\mathbb{E}_{\mathbb{P}}\left( g\left( X_{t}\right)\mid X_{t}^{i}=x^{i}\right),\quad  x^{i}\in E_{i}.
\end{equation*}
For any $i=1,2,\ldots ,m,$ the extension operator $\Phi^{i}$ acting on any function $f: E_{i}\rightarrow\mathbb{R}$ is defined by
\begin{equation*}\label{eq:PhiOperator}
\Phi^{i}f \left( x\right) =f\left( x^{i}\right),\quad  x=\left( x^{1},x^{2},\ldots ,x^{m} \right)\in E.
\end{equation*}
\end{definition}

In view of \cite[Theorem 1.11]{bjn13intricacies}, if $X$ is weakly Markovian consistent relative to $X^{i}$, then the generator matrix function of $X^{i}$, say $\Lambda_{t}^{i}$, is given as
\begin{equation}\label{eq:necCondWeak}
\Lambda_{t}^{i}= \Theta _{t}^{i}\Lambda _{t} \Phi ^{i},\quad t\geq 0.
\end{equation}
However, the converse of the above statement is not true in general. 
In Appendix \ref{apd:results}, we summarize verifiable sufficient conditions for the converse implications and several results from \cite{ysc17phd}. These results will be used later on the construction of Markov structures.
\subsection{Markov structures}\label{sec:markovStruc}

The Markov structures theory studies the stochastic dependence between the components of a multivariate Markov process, subject to marginal constraints on the coordinate processes. 
In the sequel, based on \cite{bjn13intricacies} we embark on defining \textit{Markov structures} in terms of infinitesimal generators. In analogy to strong and weak Markovian consistency, we have strong Markov structures and weak Markov structures.

\begin{definition}[Strong Markov structure]\label{def:strongMarkovStruc}
Let $\lbrace Y^{1},Y^{2},\ldots ,Y^{m}\rbrace$ be a collection of Markov chains. A multivariate process $X=(X^{1},X^{2},\ldots ,X^{m})$ is a strong Markov structure for $\lbrace Y^{1},Y^{2},\ldots ,Y^{m}\rbrace$, if
\begin{enumerate}[label=(\roman*)]
\item $X$ is a Markov chain, and $X$ satisfies the strong Markovian consistency property;
\item each component $X^{i}$ of $X$ has the same law as $Y^{i}$, $X^{i}\stackrel{\mathcal{L}}{=} Y^{i}$\footnote{The symbol $\stackrel{\mathcal{L}}{=}$ means equality in law. In the case of classical Markov chains, the initial distribution and the transition semigroup of the chain characterize the finite-dimensional distributions of the chain. Therefore, they characterize the law of the chain.}, $i=1,2,\ldots ,m$.
\end{enumerate}
\end{definition}

The counterpart of the strong Markov structure is the weak Markov structure.

\begin{definition}[Weak Markov structure]\label{def:weakMarkovStruc}
Let $\lbrace Y^{1},Y^{2},\ldots ,Y^{m}\rbrace$ be a collection of Markov chains. A multivariate process $X=(X^{1},X^{2},\ldots ,X^{m})$ is a weak Markov structure for $\lbrace Y^{1}, Y^{2},\ldots ,Y^{m}\rbrace$, if
\begin{enumerate}[label=(\roman*)]
\item $X$ is a Markov chain, and $X$ satisfies the weak Markovian consistency property;
\item each component $X^{i}$ of $X$ has the same law as $Y^{i}$, $X^{i}\stackrel{\mathcal{L}}{=} Y^{i},\ i=1,2,\ldots ,m$.
\end{enumerate}
\end{definition}

We say that a multivariate process $X$ is a \textit{weak-only Markov structure} for $\lbrace Y^{1},Y^{2},\ldots ,Y^{m}\rbrace$, if $X$ is a weak Markov structure, but it is not a strong Markov structure. Because modeling default contagion is excluded in the sense of strong Markov consistency (cf. \cite{bcch13copula, bjn13intricacies}), in practice, the weak-only Markov structure is an important class. 

One question arising naturally in the Markov structures theory is the construction of Markov structures. In view of \cite[Proposition 5.1]{bjvv08dependence}, since condition \ref{cnd:m} implies strong Markovian consistency, the construction of strong Markov structures is fully understood. Besides, we can construct infinitely many strong Markov structures for a collection of Markov chains. Detailed construction can be found in \cite{bjn13intricacies}. 

In the case of weak or weak-only Markov structures, the construction can be done by Theorem \ref{thm:semiRP} and Theorem \ref{thm:weakOnly}. In view of Theorem \ref{thm:semiRP}, the sufficient condition for the weak Markovian consistency intertwines the semigroups of $X$ and $Y^{i},\ i=1,2,\ldots ,m$. Since the semigroup of a time-inhomogeneous Markov chain is a functional of its infinitesimal generator, it turns out the construction of weak or weak-only Markov structures is not an easy task. 
We end this section by providing a continuous time algorithm for constructing weak-only Markov structures. We defer the practical viewpoint of the discrete time algorithm to Section \ref{sec:constuct}.\\

\noindent
\textbf{Algorithm I \labeltext{I}{alg:weakOnly}} (Weak-only Markov structures: continuous time)

\noindent
Input: let the initial distribution $\mu^{X}$ of $X$ and the valid generators $\left( \Lambda_{u}^{i},u\geq 0\right)$ of $Y^{i}$, $i=1,2,\ldots ,m$, be given.
\begin{enumerate}[leftmargin=*,labelindent=0pt,label= Step \arabic*.]
\item Solve for a valid infinitesimal generator $\left( \Lambda_{u},u\geq 0\right)$ of $X$, without satisfying condition \ref{cnd:m}, such that 
\begin{equation}\label{eq:intertwine}
\Theta_{t}^{i} \mathbf{P}_{t,s} = \widehat{\mathbf{P}}_{t,s}^{i} \Theta_{s}^{i} ,\quad 0\leq t\leq s,\ i=1,2,\ldots ,m,
\end{equation}
where $\widehat{\mathbf{P}}_{t,s}^{i}$ is the semigroup of $\left( \Lambda_{u}^{i},u\geq 0\right)$. Then we obtain weak Markov structures.

\item If the following condition is satisfied,
\begin{equation}\label{eq:accessible}
\mathbb{P}\left( X_{t}=\left( x^{1},x^{2},\ldots ,x^{m}\right)\right)>0,\ dt\operatorname{-a.e.} ,\quad \left( x^{1},x^{2},\ldots ,x^{m}\right)\in E,
\end{equation}
then we have the weak-only Markov structures and end the algorithm. 

If the condition \eqref{eq:accessible} is not satisfied, then we verify whether or not strong Markovian consistency property holds. If, indeed, the strong Markovian consistency does not hold, then we have the weak-only Markov structures and end the algorithm. Otherwise, we have the weak Markov structures.
\end{enumerate}

Note that for fixed $0\leq t\leq s$, \eqref{eq:intertwine} is an underdetermined homogeneous system. Moreover, the condition in Theorem \ref{thm:semiRP} is sufficient, the unavailability of the solution in the Step $1$ of Algorithm \ref{alg:weakOnly} does not imply the nonexistence of the weak Markov structure.

\section{Modeling dependence structure}\label{sec:depStru}


We consider a financial system consisting of $m$ financial institutions on a fixed finite time horizon $T>0$. We denote by $\mathcal{M}=\lbrace 1,2,\ldots ,m\rbrace$. Without loss of generality, we categorize the corporate credit rating scales to a finite state space $\mathcal{K}=\lbrace 0,1,\ldots ,K\rbrace$. By convention, the state $K$ is the default state. Suppose that $Y^{i},\ i=1,2,\ldots ,m$, are Markov chains taking values in $\mathcal{K}$. The process $Y^{i}$ represents the evolution of credit ratings of the $i$th financial institution which carries  idiosyncratic risk. The natural filtration of $Y^{i}$ can be seen as the relevant information on financial stability. We also assume that $Y^{i}$ has infinitesimal generator $\Lambda _{t}^{i} =[ \lambda_{t}^{i;x^{i},y^{i}}] _{x^{i},y^{i}\in\mathcal{K}},\ t\geq 0$, $i=1,2,\ldots ,m$, where the function $\lambda_{t}^{i;x^{i},y^{i}}$ measures the transition rate of $Y^{i}$ from rating $x^{i}$ to rating $y^{i}$ at time $t$. The transition probability function of $Y^{i}$ is denoted by $\widehat{\mathbf{P}}_{t,s}^{i}=\left[ \widehat{\mathbf{P}}_{t,s}^{i;x^{i},y^{i}}\right]_{x^{i},y^{i}\in \mathcal{K}},\ 0\leq t\leq s$, where $\widehat{\mathbf{P}}_{t,s}^{i;x^{i},y^{i}}$ represents the probability of $Y^{i}$ migrating to rating $y^{i}$ at time $s$ conditional on being in rating $x^{i}$ at time $t$.

\begin{remark}
Throughout, we assume that the marginal laws of individual credit migrating process $Y^{i}$ is known or estimated from market data. For instance, rating agencies provide transition matrices of credit ratings for financial institutions. Alternatively, many methodologies have been proposed to estimate credit rating transition matrices in literature.
\end{remark}

Next, let a continuous time multivariate Markov chain $X=\left( X^{1},X^{2},\ldots ,X^{m}\right)$ take values in $\mathcal{K}^{m}=\mathcal{K}\times \cdots \times \mathcal{K}$ with initial distribution $\mu^{X}$. The chain $X$ represents the joint credit migration process of these $m$ financial institutions. More specifically, given a pool of credit migration processes $\lbrace Y^{1},Y^{2},\ldots ,Y^{m}\rbrace$, we would like to construct the infinitesimal generator matrix  $\Lambda _{t} =\left[ \lambda _{t}^{x,y}\right] _{x,y\in\mathcal{K}^{m}},\ t\geq 0$ of $X$, such that $X$ is a Markov structure for $\lbrace Y^{1},Y^{2},\ldots ,Y^{m}\rbrace$. It follows that every component $X^{i}$ of $X$ is Markovian; necessarily, $X^{i}$ has the same probability law as $Y^{i}$,
\begin{equation}\label{eq:neceWMC}
\Theta_{t}^{i} \Lambda_{t} \Phi^{i} = \Lambda _{t}^{i},\quad t\geq 0,\ i=1,2,\ldots ,m.
\end{equation}

As already mentioned, there always exist nontrivial Markov structures. The generator $\left( \Lambda_{t}, t\geq 0\right)$ of a Markov structure is embedded with distinct algebraic structure between its components which are subject to the identical probability laws of the credit migration processes, respectively. Saying differently, the algebraic structure of every generator encodes the dependence structure between these components. We use the Markov structure as a proxy to the dependence structure of $m$ financial institutions. 

In the sequel, we dive in the details of dependence in the context of the Markov structures model.

\subsection{Independence in terms of the generator of $X$}\label{sec:dep}

We first introduce two notations. Let $\mathrm{I}$ be the identity matrix of dimension $\abs{\mathcal{K}}$. The $m$th tensor power (or the Kronecker product) of $\mathrm{I}$ is the $m$-fold tensor product of $\mathrm{I}$,
\begin{equation*}
\mathrm{I} ^{\otimes m} := \underbrace{\mathrm{I} \otimes \cdots \otimes \mathrm{I}}_{m}.
\end{equation*}
The notation $\mathrm{I} _{A}^{\otimes m,j}$ is reserved for replacing the $j$th matrix $\mathrm{I}$ by some matrix $A$ whose dimension is the same as $\mathrm{I}$,
\begin{equation*}
\mathrm{I} _{A}^{\otimes m,j} := \underbrace{\mathrm{I}\otimes \cdots \otimes \overbrace{A}^{j\text{th}} \otimes \cdots \otimes \mathrm{I}}_{m}.
\end{equation*}

An important result in \cite[Lemma 3.1 and Theorem 3.2]{bjn17note} shows that, if the semigroup of $X$ is the tensor product of the semigroups from the components, then the components of $X$ are conditional independent given $\mathcal{F}_{T}$. We apply this result here and by the fact that condition \ref{cnd:p} is equivalent to condition \ref{cnd:m}. We can show that if the infinitesimal generator of $X$ is the tensor product of the infinitesimal generators from the components, then the components of $X$ are independent. We now formally give the definition of independence between the components of $X$ by means of the infinitesimal generator of $X$.

\begin{definition}[Independence]\label{def:independence}
Let $\left( \Lambda _{t}, t\geq 0\right)$ be a valid infinitesimal generator matrix function of $X$. We say that the components of $X$ are independent if $\Lambda _{t}$ admits the representation,
\begin{equation}\label{eq:indGenerator}
\Lambda _{t} =\sum _{j=1}^{m} \mathrm{I} _{A_{t}^{j}}^{\otimes m,j} =\sum _{j=1}^{m} \underbrace{\mathrm{I}\otimes \cdots \otimes \overbrace{A_{t}^{j}}^{j\text{th}} \otimes \cdots \otimes \mathrm{I}}_{m},\quad t\geq 0,
\end{equation}
where $A_{t}^{j},\ j=1,2,\ldots ,m$, is a valid generator matrix function with the same dimension as $\mathrm{I}$.
\end{definition}

Note that an infinitesimal generator given by \eqref{eq:indGenerator} satisfies condition \ref{cnd:m}, which can be verified by computing
\begin{align*}
\Lambda _{t}\Phi^{i},\quad t\geq0,\ i=1,2,\ldots ,m.
\end{align*}
Therefore, a multivariate Markov chain having generator \eqref{eq:indGenerator} is a strong Markov structure for a collection of Markov chains generated by $\left( A_{t}^{i},t\geq 0\right),\ i=1,2,\ldots ,m$, respectively.

\subsection{Contagion}\label{sec:contagion}

As stated in Section \ref{sec:markovStruc}, we can construct infinitely many strong Markov structures by condition \ref{cnd:m}. However, a strong Markov structure satisfying condition \ref{cnd:m} has some undesirable properties in terms of risk management. In this section, we start by giving an example to explain the financial interpretations behind these properties. Meanwhile, we use this example to give intuitions about the meaning of contagion in our framework, and to elucidate how the strong Markov structure excludes contagion.

\begin{example}\label{ex:contagion}
Consider a joint credit migration process $X=\left( X^{1},X^{2}\right)$ generated by $\left( \Lambda_{t}, t\geq 0\right)$,
\begin{equation}\label{eq:tensor-contagion}
\Lambda_{t} =
\begingroup
\renewcommand*{\arraystretch}{1.2}%
\renewcommand{\kbldelim}{(}
\renewcommand{\kbrdelim}{)}
\kbordermatrix{
           & (0,0)    & (0,1)    & (1,0)   & (1,1)\cr
    (0,0)  & -(a_{t}+c_{t}+d_{t}) & d_{t}        & a_{t}         & \boldsymbol{c_{t}} \cr
    (0,1)  & f_{t}        & -(a_{t}+f_{t})  & 0             & a_{t} \cr
    (1,0)  & b_{t}        & 0              & -(b_{t}+d_{t})  & d_{t} \cr
    (1,1)  & 0        & b_{t}          & f_{t}         & -(b_{t}+f_{t}) \cr
},\endgroup
\end{equation}
where $a_{t},b_{t},c_{t},d_{t},f_{t}\geq 0$ and locally integrable.

It should be noted that in general this process $X$ is not Markovian consistent.  However, under some additional assumptions on functions $a,b,c,d,f$, the process $X$ is strongly Markovian consistent, weakly Markovian consistent, or weak-only Markovian consistent. 

Next we will argue that the strongly Markovian consistency excludes contagion between the components of $X$.

At time $t$, the infinitesimal rate of transitions for the component $X^{1}$ from state $0$ to state $1$ is
\begin{equation}\label{eq:transRate-1}
\begin{aligned}
\lambda_{t}^{(0,0),(1,0)} + \lambda_{t}^{(0,0),(1,1)} &=a_{t} +c_{t} ,&\quad \text{if } X_{t}^{2} =0,\\
\lambda_{t}^{(0,1),(1,0)} + \lambda_{t}^{(0,1),(1,1)} &=a_{t} ,&\quad \text{if } X_{t}^{2} =1.
\end{aligned}
\end{equation}
If $c_{t}=0$, then both results are the same. The state of $X_{t}^{2}$ is irrelevant. Similarly, at time $t$ the rates of jumps for $X^{1}$ from state $1$ to state $0$ are identical,
\begin{equation*}
\begin{aligned}
\lambda_{t}^{(1,0),(0,0)} + \lambda_{t}^{(1,0),(0,1)} &=b_{t},  &\quad \text{if } X_{t}^{2} =0,\\
\lambda_{t}^{(1,1),(0,0)} + \lambda_{t}^{(1,1),(0,1)} &=b_{t},  &\quad \text{if } X_{t}^{2} =1.
\end{aligned}
\end{equation*}
Thus, the infinitesimal rates of jumps for $X^{1}$ from any state to the other states do not depend on the state of $X_{t}^{2}$.

Likewise, if $c_{t}=0$, the state of $X_{t}^{1}$ does not influence the transition rate of $X^{2}$ from state $0$ to state $1$,
\begin{equation}\label{eq:transRate-2}
\begin{aligned}
\lambda_{t}^{(0,0),(0,1)} + \lambda_{t}^{(0,0),(1,1)} &=d_{t} +c_{t}, &\quad \text{if } X_{t}^{1} =0,\\
\lambda_{t}^{(1,0),(0,1)} + \lambda_{t}^{(1,0),(1,1)} &=d_{t},  &\quad \text{if } X_{t}^{1} =1,
\end{aligned}
\end{equation}
and the transition rate of $X^{2}$ from state $1$ to state $0$,
\begin{equation*}
\begin{aligned}
\lambda_{t}^{(0,1),(0,0)} + \lambda_{t}^{(0,0),(1,0)} &=f_{t}, &\quad \text{if } X_{t}^{1} =0,\\
\lambda_{t}^{(1,1),(0,0)} + \lambda_{t}^{(1,0),(1,0)} &=f_{t},  &\quad \text{if } X_{t}^{1} =1.
\end{aligned}
\end{equation*}
We say that there is no contagion between $X^{1}$ and $X^{2}$. In fact, condition \ref{cnd:m} is satisfied, thus the process $X$ is strongly Markovian consistent. This is the reason when we want to include contagion, we rule out the structures of strong Markovian consistency.

Instead, if $c_{t}>0$, given the state of $X_{t}^{2}$ the rates of jumps for $X^{1}$ from state $0$ to state $1$ in \eqref{eq:transRate-1} are different. The state of $X_{t}^{2}$ will have an impact on how $X^{1}$ jumps from state $0$ to state $1$. Similarly, the state of $X_{t}^{1}$ affects the rate of transitions for $X^{2}$ from state $0$ to state $1$ in \eqref{eq:transRate-2}. Then we say there exists contagion between $X^{1}$ and $X^{2}$.
\end{example}

\begin{remark}
Note that in general the entry of simultaneous jump with zero value, for example $c_{t}=0$, does not imply that the components cannot jump simultaneously, or the probability of the common jump is zero. The probability of the simultaneous jump depends on the structure of $\Lambda_{t}$. We need to check the corresponding transition probability matrix. Actually, given functions $a,b,d,f>0$ and $c=0$ in \eqref{eq:tensor-contagion} the probabilities of all simultaneous jumps are nonzero.
\end{remark}

Another financial perspective of condition \ref{cnd:m} is the \textit{financial resilience}. In Example \ref{ex:contagion}, if $c_{t}=0$, the default of either party has no impact on the default of the counterparty. This does not mean that $X^{1}$ and $X^{2}$ are independent or unconnected. Instead, it implies that the corresponding financial institutions of $X^{1}$ and $X^{2}$ are financially resilient to the default of the counterparty. It is an important feature in our model that we allow the flexibility for some financial institutions not to be affected by the defaults of some financial institutions.

\subsection{Construction of the weak-only Markov structures}\label{sec:constuct}

Since strong Markovian consistency implies weak Markovian consistency, we can construct a larger class of Markov structures, namely, weak-only Markov structures, to explore various dependence structures between financial institutions. In view of Theorem \ref{thm:semiRP}, the sufficient conditions for weak Markovian consistency are formulated by the intertwining the semigroups of $X$ and the prescribed marginal processes. In the sequel, we assume that \textit{all} financial institutions are affected by the default of \textit{any} financial institution. We discuss how to construct weak-only Markov structures for different types of input data. \\

\noindent
\textbf{Given continuous data $\Lambda_{u}^{i},\ 0\leq u\leq T,\ i=1,2,\ldots ,m$}\label{sec:contModel}

\noindent
Now, suppose that the model examiner predicts the marginal probability laws from today to some future time $T$, i.e., $\left( \Lambda_{u}^{i},0\leq u\leq T\right),\ i=1,2,\ldots ,m$. If the goal is to construct continuous time weak-only Markov structures between time $0$ to time $T$, the Algorithm \ref{alg:weakOnly} in Section \ref{sec:markovStruc} provides one way to construct the weak-only Markov structures starting from today to time $T$. Namely, \textit{today} will be the initial time $0$. Since the identity \eqref{eq:intertwine} is required to be satisfied for any $0\leq t\leq s\leq T$, seeking the algebraic relations between the entries of $\Lambda_{u},\ 0\leq u\leq T$ is less likely. Under this circumstance, the structure imposed on the generator matrix $\left( \Lambda_{u},0\leq u\leq T\right)$ becomes crucial. 
Some examples are presented in Section \ref{sec:numerical}.\\

\noindent
\textbf{Given discrete data $\Lambda_{{t_{n+1}}}^{i},\ n\in\mathbb{N},\ i=1,2,\ldots ,m,$}\label{sec:discModel}

\noindent
Alternatively, assume that the input marginal generators are provided one step ahead of time. Saying differently, suppose that today is at time $t=t_{n},\ n\in\mathbb{N}$. The input data is $\Lambda_{t_{n+1}}^{i},\ i=1,2,\ldots ,m$. In this case, the objective is to construct the weak-only Markov structures from time $t_{n}$ to time $t_{n+1}$, step by step. Since we model the joint credit migration process as a Markov structure, by the virtue of Markov property, whatever happened in the past does not affect the future. Instead of the initial distribution of $X$, we care about the one-dimensional distribution of $X$ at every single time. We modify Algorithm \ref{alg:weakOnly} by the following.\\


\noindent
\textbf{Algorithm II \labeltext{II}{alg:weakOnlyDisc}}(Weak-only Markov structures: discrete time)

\noindent
Input: let valid generators $\Lambda_{t_{n+1}}^{i}$ of $Y^{i}$, $i=1,2,\ldots ,m$, and a non-trivial distribution of $X_{t_{n}}$ be given, i.e., 
\begin{equation}\label{eq:accessible-1}
\mathbb{P}\left( X_{t_{n}}=\left( x^{1},x^{2},\ldots ,x^{m}\right)\right)>0,\quad \left( x^{1},x^{2},\ldots ,x^{m}\right)\in \mathcal{K}^{m}.
\end{equation}

\begin{enumerate}[leftmargin=*,labelindent=0pt,label= Step \arabic*.]
\item Solve for valid infinitesimal generators $\Lambda_{t_{n+1}}$ without satisfying condition \ref{cnd:m} such that 
\begin{equation}\label{eq:intertwine-1}
\Theta_{t_{n}}^{i} \mathbf{P}_{t_{n},t_{n+1}} = \widehat{\mathbf{P}}_{t_{n},t_{n+1}}^{i} \Theta_{t_{n+1}}^{i} ,\quad i=1,2,\ldots ,m,\ n\in\mathbb{N},
\end{equation}
where $\widehat{\mathbf{P}}_{t_{n},t_{n+1}}^{i}$ is the semigroup of $\Lambda_{t_{n+1}}^{i}$, $i=1,2,\ldots ,m$.

\item If the solution $\Lambda_{t_n+1}$ exists, then we have the weak-only Markov structures up to time $t_{n+1}$ and move forward to next step. Otherwise, the weak-only Markov structure does not exist with the given one-dimensional distribution $\mathbb{P}\left( X_{t_{n}}=\left( x^{1},x^{2},\ldots ,x^{m}\right)\right)$.

\item We run the algorithm recursively until the terminal time $T$ and end the algorithm.
\end{enumerate}

Note that, if we fix the same ordering of the state spaces $\mathcal{K}$ and $\mathcal{K}^{m}$ for every operator $\Phi^{i}$ and $\Theta_{t_{n}}^{i}$, the homogeneous system \eqref{eq:intertwine-1} is underdetermined. Specifically, we have $\left( \abs{\mathcal{K}^{m}}-1\right)\times \abs{\mathcal{K}^{m}}$ unknowns $\lambda_{t_{n+1}}^{x,y}$ and $m\times\abs{\mathcal{K}}\times \left( \abs{\mathcal{K}}-1\right)\times \abs{\mathcal{K}}$ equations. It holds that 
\begin{equation*}
\abs{\mathcal{K}^{m}} \left( \abs{\mathcal{K}^{m}}-1\right)\geq m \abs{\mathcal{K}}^{2}\left( \abs{\mathcal{K}}-1\right),\quad m,\abs{\mathcal{K}}\geq 2 .
\end{equation*}
In general, such a solution $\Lambda_{t_{n+1}}$ is not unique. If the solution $\Lambda_{t_{n+1}}$ is identified, then we construct $ \Lambda_{t_{n+1}}$ forward and one step at a time, as the input marginals $\Lambda_{t_{n+1}}^{i},\ i=1,2,\ldots ,m$, become available. 

As time step approaches the infinitesimal level, in light of next proposition, we argue that \eqref{eq:intertwine-1} remains to hold for any larger time steps. The constructed Markov structure in a forward recursive approach is indeed a Markov structure up to time $T$.

\begin{proposition}\label{prop:semiDelta}
Let the assumptions in Theorem \ref{thm:semiRP} hold true. If we have 
\begin{equation*}
\begin{split}
\Theta_{t}^{i} \mathbf{P}_{t,v} = \widehat{\mathbf{P}}_{t,v}^{i} \Theta_{v}^{i},
\end{split}
\qquad\text{and}\qquad
\begin{split}
\Theta_{v}^{i} \mathbf{P}_{v,s} = \widehat{\mathbf{P}}_{v,s}^{i} \Theta_{s}^{i},\quad 0\leq t\leq v\leq s,
\end{split}
\end{equation*}
then
\begin{equation*}
\Theta_{t}^{i} \mathbf{P}_{t,s} = \widehat{\mathbf{P}}_{t,s}^{i} \Theta_{s}^{i}.
\end{equation*}
\end{proposition}

\begin{proof}
Since $\mathbf{P}_{t,s}$ and $\widehat{\mathbf{P}}_{t,s}^{i}$ satisfy the semigroup property, the result follows
\begin{equation*}
\Theta_{t}^{i} \mathbf{P}_{t,s} = \Theta_{t}^{i} \mathbf{P}_{t,v} \mathbf{P}_{v,s} = \widehat{\mathbf{P}}_{t,v}^{i} \Theta_{v}^{i} \mathbf{P}_{v,s} = \widehat{\mathbf{P}}_{t,v}^{i} \widehat{\mathbf{P}}_{v,s}^{i} \Theta_{s}^{i} = \widehat{\mathbf{P}}_{t,s}^{i} \Theta_{s}^{i}.
\end{equation*}
\end{proof}





\subsection{Calibration of the dependence structure}\label{sec:calibration}

As for the dependence structure, in case market prices of any ideal basket credit products are available, we can derive theoretical prices and calibrate the dependence structure through market prices. However, restricted to the current situation, it is rather unlikely. From the practical point of view, one possible procedure may simulate several Markov structures and target the probability distribution of $X_{t_{n+1}}$, i.e., $\mathbb{P}\left( X_{t_{n+1}} =\left( x^{1},x^{2},\ldots ,x^{m}\right)\right),\ \left( x^{1},x^{2},\ldots ,x^{m}\right)\in \mathcal{K}^{m}$ at every time step. Each Markov structure for the given collection of Markov chains stands for a possible scenario of dependence structure of the financial economy. Once the stylized Markov structure is identified, the rest Markov structures can be used in a stress test. 

\section{Measuring systemic risk, systemic dependence, and systemic instability}\label{sec:sr}

When more than one Markov structure exists, i.e., the dependent Markov structures, there is a potential uncertainty caused by the dependence structures. In what follows, we construct the systemic risk measure, the systemic dependence measure, and the systemic instability measure to explore this matter.

Recall that $\lbrace Y^{1},\ldots ,Y^{m}\rbrace$ is a collection of Markov chains. Every process $Y^{i}$ has generator $\left( \Lambda_{u}^{i}, u\geq 0\right)$ with initial distribution $\mu^{Y^{i}}$, $i=1,2,\ldots ,m$, respectively. We construct a valid infinitesimal generator $\left( \Lambda^{\mathcal{I}} _{u},u\geq 0\right)$ by
\begin{equation*}
\Lambda^{\mathcal{I}} _{u} =\sum _{j=1}^{m} \mathrm{I} _{\Lambda_{u}^{j}}^{\otimes m,j}, \quad u\geq 0.
\end{equation*}
As known from Definition \ref{def:independence}, a joint credit migration process modeled by $\left( \Lambda_{u}^{\mathcal{I}},u\geq 0\right)$ has independent components. We say this joint credit migration process has independence structure. We denote by $\mathcal{I}$ the independence structure and call this financial system with fully independent financial institutions as the \textit{neutral} financial system. 

In view of the Kolmogorov Existence Theorem, on the same probability space $\left( \Omega ,\mathcal{F},\mathbb{P}\right)$ there exists a unique multivariate Markov chain, say  $X^{\mathcal{I}}=(X^{\mathcal{I},1},\ldots ,X^{\mathcal{I},m})$, generated by $\left( \Lambda^{\mathcal{I}} _{u},u\geq 0\right)$. Since $\left( \Lambda^{\mathcal{I}} _{u},u\geq 0\right)$ satisfies condition \ref{cnd:m}, $X^{\mathcal{I}}$ is strongly Markovian consistent. Naturally, the component $X^{\mathcal{I},i}$ is generated by $\left( \Theta_{u}^{i}\Lambda_{u}^{\mathcal{I}} \Phi^{i}, u\geq 0\right)$. If we assume that the initial distribution $\mu^{X^{\mathcal{I},i}}$ is identical to $\mu^{Y^{i}}$, then every $X^{\mathcal{I},i}$ has the same probability law as $Y^{i}$. It turns out $X^{\mathcal{I}}$ is a strong Markov structure. We would like to emphasize that contagion is excluded from a joint credit migration process modeled by $X^{\mathcal{I}}$.

Now, we denote by $z_{i}\in\mathcal{K},\ i=1,,\ldots ,m$, corresponding to some credit rating. Let $z_{i}$, $0\leq t\leq T<\infty$, and $h\in\mathcal{M}$ be fixed. We would like to compute the conditional probability that, at future time $T$, at least $h$ financial institutions are in credit rating $z_{i}$ conditional on the information available at time $t$. That is,
\begin{equation}\label{eq:srm-ind}
\mathbb{P}\left( \sum _{i=1}^{m} \mathbbm{1}_{\lbrace X_{T}^{{\mathcal{I}},i}=z_{i}\rbrace} \geq h\ \bigg\rvert\ X_{t}^{\mathcal{I}} \right)  ,\quad z_{i}\in\mathcal{K},\ h\in\mathcal{M},\ 0\leq t\leq T.
\end{equation}

Suppose that we model the joint credit migration process $X$ as a Markov structure for the collection $\lbrace Y^{1},\ldots ,Y^{m}\rbrace$. Then the components of $X$ are Markovian, and necessarily generated by
\begin{equation*}
\Lambda_{u}^{i} =\Theta_{u}^{i} \Lambda_{u} \Phi^{i},\quad u\geq 0,\ i=1,2,\ldots ,m.
\end{equation*}
If a nontrivial solution $\left( \Lambda^{\mathcal{D}}_{u},u\geq 0\right)$ exists, which is a solution different from $\left( \Lambda^{\mathcal{I}}_{u},u\geq 0\right)$, it defines a dependence structure, say $\mathcal{D}$. Accordingly, we compute the conditional probability
\begin{equation}\label{eq:srm-dep}
\mathbb{P}\left( \sum _{i=1}^{m} \mathbbm{1}_{\left\lbrace X_{T}^{\mathcal{D},i}=z_{i}\right\rbrace} \geq h\ \bigg\rvert\  X_{t}^{\mathcal{D}} \right) ,\quad z_{i}\in\mathcal{K},\ h\in\mathcal{M},\ 0\leq t\leq T,
\end{equation}
where $X^{\mathcal{D}}$ has generator $\left( \Lambda^{\mathcal{D}}_{u},u\geq 0\right)$. If $X^{\mathcal{D}}$ is weak-only Markovian consistent, a joint credit migration process modeled by $X^{\mathcal{D}}$ allows contagion between individual credit migrations.

Before we introduce the following definitions, let $\mathcal{X}$ denote the collection of Markov chains with values in $\mathcal{K}^m$. We give a definition for systemic risk measure.

\begin{definition}[Systemic Risk Measure]\label{def:srm}
Suppose that $X^{\mathcal{D}}$ is a Markov structure for a collection of Markov chains $\lbrace Y^{1},Y^{2},\ldots ,Y^{m}\rbrace$. Let $z=\left( z_{1},z_{2},\ldots ,z_{m}\right)\in\mathcal{K}^{m}$ be fixed. We define a function $\nu^{z}: [0,\infty )\times [0,T]\times \mathcal{M}\times \mathcal{X} \times \mathcal{K}^m \rightarrow [0,1]$,
\begin{equation*}\label{eq:srmRealized}
\nu^{z} \left( T,t,h,X^{\mathcal{D}},x\right) :=\mathbb{P}\left( \sum _{i=1}^{m} \mathbbm{1}_{\left\lbrace X_{T}^{\mathcal{D},i}=z_{i}\right\rbrace} \geq h\ \bigg\rvert\ X^{\mathcal{D}}_{t}=x \right) .
\end{equation*}
In particular, if every $z_{i}$ is the corresponding default rating, we call this function \textbf{the systemic risk measure}.
\end{definition}


In view of Definition \ref{def:srm}, if we ignore the dependence structure, the concept of measuring systemic risk is quite common in finance literature, for instance, the joint default probability of $m$ financial institutions. We argue that, using this systemic risk measure to monitor systemic risk is not rigorous. Because this measure neglects the dependence structure within the financial system and solely signals systemic risk once the probability reaches predetermined level. It may happen that different dependence structures contribute to the same level of systemic risk. 

Accordingly, we define the dependence measure as the difference between \eqref{eq:srm-ind} and \eqref{eq:srm-dep}. Saying differently, we normalize the probability obtained from the dependence structure with respect to the independence structure. Since the Markov structures $X^{\mathcal{D}}$ and $X^{\mathcal{I}}$ are subject to the same pool of marginal laws, the normalization of \eqref{eq:srm-dep} has solid grounds. When every $z_{i}$ is the corresponding default rating, we have the systemic dependence measure.

\begin{definition}[Systemic Dependence Measure]\label{def:sdm}
Suppose that $X^{\mathcal{D}}$ is a Markov structure for a collection of Markov chains $\lbrace Y^{1},Y^{2},\ldots ,Y^{m}\rbrace$. Let $z=\left( z_{1},z_{2},\ldots ,z_{m}\right)\in\mathcal{K}^{m}$ be fixed. We define a function $\rho^{z}: [0,\infty )\times [0,T]\times \mathcal{M}\times \mathcal{X} \times \mathcal{K}^{m} \rightarrow [-1,1]$,
\begin{equation}\label{eq:sdmRealized}
\rho^{z} \left( T,t,h,X^{\mathcal{D}},x\right) :=\nu^{z} \left( T,t,h,X^{\mathcal{D}},x\right) - \nu^{z} \left( T,t,h,X^{\mathcal{I}},x\right).
\end{equation}
In particular, if every $z_{i}$ is the corresponding default rating, we call this function \textbf{the systemic dependence measure}.
\end{definition}

In view of Definition \ref{def:sdm}, $\rho^{z} \left( T,t,h,X^{\mathcal{D}},x\right)$ measures the disequilibrium with respect to the neutral system, i.e., the independence structure. Not only do we compute the conditional probability of at least $h$ financial institutions being in default at time $T$, but we also concern this probability deviating from the neutral system to encompass the effect resulting from the dependence structure. 

Note that, we restrict ourselves to the Markov structure $X^{\mathcal{D}}$ starting from the same initial distribution as $X^{\mathcal{I}}$.  
Although processes $X^{\mathcal{D}}$ and $X^{\mathcal{I}}$ evolve from the same initial distribution, their one-dimensional distributions at time $t$, $\mathbb{P}\left( X_{t}^{\mathcal{D}} =x\right)$ and $\mathbb{P}\left( X_{t}^{\mathcal{I}} =x\right)$, are generally different. In the following, we prove that the distribution of $X_{t}^{\mathcal{D}}$ is absolutely continuous with respect to the distribution of $X_{t}^{\mathcal{I}}$.

\begin{proposition}
Assume that $X^{\mathcal{I}}$ and $X^{\mathcal{D}}$ are Markov structures for a collection of Markov chains $\lbrace Y^{1},Y^{2},\ldots ,Y^{m}\rbrace$. Suppose that $X^{\mathcal{I}}$ and $X^{\mathcal{D}}$ evolve from the same initial distribution. Then the distribution of $X_{t}^{\mathcal{D}}$ is absolutely continuous with respect to the distribution of $X_{t}^{\mathcal{I}}$.
\end{proposition}

\begin{proof}
Let $x = \left( x_{1},x_{2},\ldots ,x_{m}\right)\in\mathcal{K}^{m}$. Assume that $\mathbb{P}\left( X_{t}^{\mathcal{I}} = x\right) = 0$ for some $t$,  
\begin{equation*}
\mathbb{P}\left( X_{t}^{\mathcal{I}} = x\right) = \sum_{z\in\mathcal{K}^{m}} \mathbb{P}\left( X_{0}^{\mathcal{I}} = z\right) \mathbb{P}\left( X_{t}^{\mathcal{I}} = x\mid X_{0}^{\mathcal{I}} = z\right) = 0.
\end{equation*}
It remains to show $\mathbb{P}\left( X_{t}^{\mathcal{D}} = x\right) =0$. Indeed, if the initial distribution is nontrivial, i.e., $\mathbb{P}\left( X_{0}^{\mathcal{I}} = z\right) \neq 0$ for some $z$, then $\mathbb{P}\left( X_{t}^{\mathcal{I}} = x\mid X_{0}^{\mathcal{I}} = z\right) =0$. By independence of $X^{\mathcal{I}}$, we know at least one of $\mathbb{P}\left( X_{t}^{\mathcal{I},i} = x_{i}\mid X_{0}^{\mathcal{I},i} = z_{i}\right) =0,\ i= 1,2,\ldots ,m$. Since $X^{\mathcal{I},i}$ and $X^{\mathcal{D},i}$ have the identical law as $Y^{i}$, 
it follows immediately $\mathbb{P}\left( X_{t}^{\mathcal{D},i} = x_{i}\mid X_{0}^{\mathcal{D},i} = z_{i}\right) =0$ for the same component of $X^{\mathcal{I}}$ with $\mathbb{P}\left( X_{t}^{\mathcal{I},i} = x_{i}\mid X_{0}^{\mathcal{I},i} = z_{i}\right) =0$.

Clearly, because $X_{t}^{\mathcal{D}}$ is a multivariate Markov chain, we must have $\mathbb{P}\left( X_{t}^{\mathcal{D}} = x\mid X_{0}^{\mathcal{D}} = z\right) =0$. Thus, for the same nontrivial initial distribution,
\begin{equation*}
\mathbb{P}\left( X_{t}^{\mathcal{D}} = x\right) =\sum_{z\in\mathcal{K}^{m}} \mathbb{P}\left( X_{0}^{\mathcal{D}} = z\right) \mathbb{P}\left( X_{t}^{\mathcal{D}} = x\mid X_{0}^{\mathcal{D}} = z\right) = 0.
\end{equation*}
\end{proof}

To this end, motivated by L\'{o}pez-Ruiz, Mancici and Calbet \cite{RMC2010scm}, in order to emphasize the uncertainty resulting from the distributions of $X_{t}^{\mathcal{D}}$ and $X_{t}^{\mathcal{I}}$, we introduce the instability measure. In what follows, we adopt the convention $\log \left( \frac{0}{0}\right) = 0$.

\begin{definition}[Systemic Instability Measure]\label{def:sim}
Suppose that $X^{\mathcal{D}}$ is a Markov structure for a collection of Markov chains $\lbrace Y^{1},Y^{2},\ldots ,Y^{m}\rbrace$. Let $z=\left( z_{1},z_{2},\ldots ,z_{m}\right)\in\mathcal{K}^{m}$ be fixed. We define a bounded function $\kappa^{z}$ from $[0,\infty )\times [0,T]\times \mathcal{M}\times \mathcal{X} \times \mathcal{K}^{m} \rightarrow\mathbb{R}$,
\begin{equation}\label{eq:scmRealized}
\kappa^{z} \left( T,t,h,X^{\mathcal{D}},x\right) :=\rho^{z} \left( T,t,h,X^{\mathcal{D}},x\right) \times \left( \sum_{y\in\mathcal{K}^{m}} \mathbb{P}\left( X_{t}^{\mathcal{D}} = y\right) \log \frac{\mathbb{P}\left( X_{t}^{\mathcal{D}} = y\right)}{\mathbb{P}\left( X_{t}^{\mathcal{I}} = y\right)}\right).
\end{equation}
In particular, if every $z_{i}$ is the corresponding default rating, we call this function \textbf{the systemic instability measure}.
\end{definition}

It should be noted that the second term in \eqref{eq:scmRealized} measures the \textit{distance} between the one-dimensional distributions of $X^{\mathcal{D}}$ and $X^{\mathcal{I}}$ at time $t$. This quantity is also known as Kullback-Leibler divergence or relative entropy. In our case, the term is actually the mutual information $I\left( X_{t}^{\mathcal{D},1}, X_{t}^{\mathcal{D},2},\ldots , X_{t}^{\mathcal{D},m}\right)$. Due to the non-negativity of the relative entropy, this term will not flip the sign of the systemic dependence measure. Instead, the second term in \eqref{eq:scmRealized} will scale the dependence measure based on their one-dimensional distributions. The properties of the Kullback-Leibler divergence are well-studied, for instance, the upper bounds. We refer the reader to \cite{cover06eit, pd16bounds}, and references therein. 

\subsection{Classifications of systemic dependence measure and systemic instability measure}\label{sec:sdmType}

We have created measures to monitor financial stability. From now we denote by $K$ the default rating and we focus on $\left( \nu^{z}, \rho^{z}, \kappa^{z}\right)$ for $z=\left(K,\ldots ,K\right)$. 

In view of Definition \ref{def:sdm}, the systemic dependence measure depends on $T,t,h$ and $\mathcal{D}$. 
It should be noted that the systemic dependence measure is evaluated based on different dependence structures. For instance, let $X^{\mathcal{D}}$ be a weak-only Markov structure. Then we can compare the conditional probability of the event under a joint credit migration process \textit{embedded} with contagious mechanism to the conditional probability of the same event under a joint credit migration process \textit{excluding} contagion.

Financially speaking, when we fix $h=m$ and $T$, the event of interest is all financial institutions in default at some future time $T$. We measure the probability of this financial system that will be in \textit{Armageddon} at future time $T$. 
It follows that the systemic dependence measure depends on $t,X_{t}$. In view of \eqref{eq:sdmRealized}, at time $t$ and $X_{t}^{\mathcal{D}}=x$, if the systemic dependence measure $\rho^{z} \left( T,t,h,X^{\mathcal{D}},x\right)>0$, it means that the probability of all financial institutions in default rating at future time $T$ under a joint credit migration process with dependence structure $\mathcal{D}$ is strictly larger than the probability of the same event under the process with independence structure. In other words, the financial system with dependence structure $\mathcal{D}$ has greater exposure to the risk in Armageddon. Under this dependence structure $\mathcal{D}$, the financial system is no better than the neutral financial system. We say for the pair $(t,x)$, this financial system with dependence structure $\mathcal{D}$ is of \textit{unfavorable systemic dependence}.

At time $t$ and $X_{t}^{\mathcal{D}}=x$, if the systemic dependence measure $\rho^{z} \left( T,t,h,X^{\mathcal{D}},x\right)=0$, it means that the financial system with dependence structure $\mathcal{D}$ behaves like the neutral financial system. Then we call this financial system with dependence structure $\mathcal{D}$ a \textit{neutral systemic dependence} for the pair $(t,x)$. By analogy, if at time $t$ and $X_{t}^{\mathcal{D}}=x$ the systemic dependence measure $\rho^{z} \left( T,t,h,X^{\mathcal{D}},x\right) <0$, it suggests that the financial system with dependence structure $\mathcal{D}$ has less exposure to the risk of simultaneously default at future time $T$. This financial system benefits from the current dependence structure. We say that for the pair $(t,x)$ this financial system with dependence structure $\mathcal{D}$ has \textit{favorable systemic dependence}.

According to the sign of this systemic dependence measure, we categorize the dependence status of the financial system into the following:
\begin{align*}
\rho^{z} \left( T,t,m,X^{\mathcal{D}},x \right) \begin{cases}
>0,\quad \text{unfavorable systemic dependence};\\
=0,\quad \text{neutral systemic dependence};\\
<0,\quad \text{favorable systemic dependence}.
\end{cases}
\end{align*}
Since the Kullback-Leibler divergence will not change the sign of the systemic dependence measure, the systemic instability measure is classified as three cases:
\begin{align*}
\kappa^{z} \left( T,t,m,X^{\mathcal{D}},x \right) \begin{cases}
>0,\quad \text{systemic risk};\\
=0,\quad \text{systemic indifference};\\
<0,\quad \text{systemic benefit}.
\end{cases}
\end{align*}

\subsection{Properties of systemic dependence measure and systemic instability measure}\label{sec:scdmProperty}

In this section, we provide two important properties of our systemic dependence measure and systemic instability measure. These properties are mainly inherited from the virtues of the Markov structures theory. The first property is that both measures are invariant with respect to the permutation of the components. This property is essential. No matter how the ordering of the financial institutions is, the ordering will not change the values of these two systemic measures. The second property is the law invariance. The major financial meaning is that subject to the same pool of financial institutions, only the financial systems with the identical dependence structure will produce the same levels of the systemic dependence measure and systemic instability measure at every time.

We start with the important properties of Kronecker product. 

\begin{lemma}
\begin{enumerate}[label =(\roman*)]
\item (\cite[4.2.6]{horn84matrix}) Let $A,B,C$ be matrices. Then $\left( A\otimes B\right) \otimes C = A\otimes \left( B \otimes C\right)$.

\item (\cite[Section 3]{hs81vec}) Let $A$ and $B$ be two matrices. Then $A\otimes B$ and $B\otimes A$ are permutation equivalent. Namely, there exists permutation matrices $P,Q$ such that
\begin{equation*}
A\otimes B = P \left( B\otimes A\right) Q .
\end{equation*}
In particular, if $A$ and $B$ are square matrices, then $A\otimes B$ and $B\otimes A$ are permutation similar. We can take $P=Q^{\top}$.
\end{enumerate}
\end{lemma}

Next, we show that although rearranging the components of $X$ will change the matrix defined in the right hand side of \eqref{eq:indGenerator}, the independence relationship between the components will not be altered.

\begin{proposition}\label{prop:perSimilar}
The infinitesimal generator constructed by \eqref{eq:indGenerator} is permutation similar with respect to the components of $X = \left( X^{1},X^{2},\ldots ,X^{m}\right)$. 
\end{proposition}

\begin{proof}
We prove by induction. It is clear that for $m=1$ any matrix is similar to itself. Let $m=2$. We fix the ordering of state space of $X=\left( X^{1},X^{2}\right)$. Assume that the permutation $\pi$ with respect to the state space is given by
\begin{align}
\pi = 
\begingroup
\renewcommand*{\arraystretch}{1.2}%
\renewcommand{\kbldelim}{(}
\renewcommand{\kbrdelim}{)}
\kbordermatrix{
           &     &    &    & \cr
    & (0,0)  & (0,1)   & \cdots   & (K,K) \cr
    & \pi \left( (0,0)\right) & \pi \left( (0,1)\right) & \cdots & \pi \left( (K,K)\right) \cr
}.\endgroup \label{eq:perState}
\end{align}
The permutation matrix $P_{\pi}$ corresponding to the permutation $\pi$ is of the form,
\begin{align}
P_{\pi} = 
\begingroup
\renewcommand*{\arraystretch}{1.2}%
\renewcommand{\kbldelim}{(}
\renewcommand{\kbrdelim}{)}
\kbordermatrix{
           &    \cr  
    & e_{\pi \left( (0,0)\right)} \cr
    & e_{\pi \left( (0,1)\right)} \cr
    & \vdots \cr 
    & e_{\pi \left( (K,K)\right)} \cr
},\endgroup \label{eq:perMatrix}
\end{align}
where $e_{j}$ is the canonical basis. Then $\Lambda_{t}$ is permutation similar to $A_{t}^{2} \otimes \mathrm{I} + \mathrm{I} \otimes A_{t}^{1}$,
\begin{align*}
\Lambda_{t} &= A_{t}^{1} \otimes \mathrm{I} + \mathrm{I} \otimes A_{t}^{2} \\
 &= P_{\pi} \left( \mathrm{I} \otimes A_{t}^{1} \right) P_{\pi}^{\top}  + P_{\pi} \left( A_{t}^{2} \otimes \mathrm{I}\right) P_{\pi}^{\top} \\
 &= P_{\pi} \left( A_{t}^{2} \otimes \mathrm{I} + \mathrm{I} \otimes A_{t}^{1}\right) P_{\pi}^{\top} .
\end{align*}
In view of Definition \ref{def:independence}, a process generated by the infinitesimal generator $A_{t}^{2} \otimes \mathrm{I} + \mathrm{I} \otimes A_{t}^{1}$ has components ordered by $\left( X^{2},X^{1}\right)$. We can prove by induction on the number of components of $X$ for $m\geq 3$.
\end{proof}

\begin{remark}
Let $X^{1}$ and $X^{2}$ be distinct. The independence generators for the processes with the components ordered by $\left( X^{1},X^{2}\right)$ and by $\left( X^{2},X^{1}\right)$ are different. But they contain the same information up to permutation matrices.
\end{remark}


\begin{proposition}\label{prop:perSimilarComponent}
A Markov structure $X=\left( X^{1}, X^{2},\ldots ,X^{m}\right)$ for a collection of Markov chains $\lbrace Y^{1}, Y^{2},\ldots ,Y^{m}\rbrace$ is permutation similar with respect to the permutation of $\lbrace Y^{1}, Y^{2},\ldots ,Y^{m}\rbrace$.
\end{proposition}

\begin{proof}
When we choose the ordering of the state space of $Y^{i}$ and the arrangement of the components of this collection, the state space of $X$ and $X^{i}$ will be fixed. Now, we permute $\lbrace Y^{1}, Y^{2},\ldots ,Y^{m}\rbrace$ and the state space of $Y^{i}$'s. Let $P_{\pi}$ be the permutation matrix corresponding to the permutation $\pi$ with respect to the state space of $X$. We denote by $P_{\overbar{\pi}}$ the matrix representation of the permutation $\overbar{\pi}$ with respect to the state space of $X^{i}$,
\begin{align}
\overbar{\pi} = 
\begingroup
\renewcommand*{\arraystretch}{1.2}%
\renewcommand{\kbldelim}{(}
\renewcommand{\kbrdelim}{)}
\kbordermatrix{
           &     &    &    & \cr
    & 0  & 1   & \cdots   & K \cr
    & \overbar{\pi} \left( 0\right) & \overbar{\pi} \left( 1\right) & \cdots & \overbar{\pi} \left( K\right) \cr
}.\endgroup \label{eq:perComponent}
\end{align}
Since the transition probability matrix is non-singular, we can always perform similarity transformation for any transition probability matrix. We have that $\mathbf{P}_{t,s}$ and $\mathbf{P}_{t,s}^{\pi(i)}$ are permutation-similar, so are $\widehat{\mathbf{P}}_{t,s}^{i}$ and $\widehat{\mathbf{P}}_{t,s}^{\overbar{\pi}(i)}$,
\begin{equation*}
\mathbf{P}_{t,s} = P_{\pi} \mathbf{P}_{t,s}^{\pi(i)} P_{\pi}^{\top},
\end{equation*}
and
\begin{equation*}
\widehat{\mathbf{P}}_{t,s}^{i} = P_{\overbar{\pi}} \widehat{\mathbf{P}}_{t,s}^{\overbar{\pi}(i)} P_{\overbar{\pi}}^{\top}.
\end{equation*}

Note that $\Theta_{t}^{i}\mathbf{P}_{t,s} = \widehat{\mathbf{P}}_{t,s}^{i} \Theta_{s}^{i},\ i=1,2,\ldots ,m$, is the sufficient condition for weak Markovian consistency. It suffices to show this sufficient condition remains true under the permutation of the components. 
Replacing the similarity transformation of $\mathbf{P}_{t,s}$ and $\widehat{\mathbf{P}}_{t,s}^{i}$ into the sufficient condition, it follows that
\begin{align*}
\Theta_{t}^{i} P_{\pi} \mathbf{P}_{t,s}^{\pi(i)} P_{\pi}^{\top} &= P_{\overbar{\pi}} \widehat{\mathbf{P}}_{t,s}^{\overbar{\pi}(i)} P_{\overbar{\pi}}^{\top} \Theta_{s}^{i} \\
\Theta_{t}^{i} P_{\pi} \mathbf{P}_{t,s}^{\pi(i)} P_{\pi}^{\top} P_{\pi} &= P_{\overbar{\pi}} \widehat{\mathbf{P}}_{t,s}^{\overbar{\pi}(i)} P_{\overbar{\pi}}^{\top} \Theta_{s}^{i} P_{\pi} \\
\Theta_{t}^{i} P_{\pi} \mathbf{P}_{t,s}^{\pi(i)} &= P_{\overbar{\pi}} \widehat{\mathbf{P}}_{t,s}^{\overbar{\pi}(i)} P_{\overbar{\pi}}^{\top} \Theta_{s}^{i} P_{\pi} \\
P_{\overbar{\pi}}^{\top} \Theta_{t}^{i} P_{\pi} \mathbf{P}_{t,s}^{\pi(i)} &= P_{\overbar{\pi}}^{\top} P_{\overbar{\pi}} \widehat{\mathbf{P}}_{t,s}^{\overbar{\pi}(i)} P_{\overbar{\pi}}^{\top} \Theta_{s}^{i} P_{\pi} \\
P_{\overbar{\pi}}^{\top} \Theta_{t}^{i} P_{\pi} \mathbf{P}_{t,s}^{\pi(i)} &= \widehat{\mathbf{P}}_{t,s}^{\overbar{\pi}(i)} P_{\overbar{\pi}}^{\top} \Theta_{s}^{i} P_{\pi} \\
\widetilde{\Theta}_{t}^{\pi(i)} \mathbf{P}_{t,s}^{\pi(i)} &= \widehat{\mathbf{P}}_{t,s}^{\overbar{\pi}(i)} \widetilde{\Theta}_{s}^{\pi(i)},
\end{align*}
where $\widetilde{\Theta}_{t}^{\pi(i)} := P_{\overbar{\pi}}^{\top} \Theta_{t}^{i} P_{\pi}$, and for any permutation matrix $P_{\pi}^{-1} = P_{\pi}^{\top}$. 

Since $P_{\overbar{\pi}}^{\top} \Theta_{t}^{i} P_{\pi}$ merely changes the orders of columns and rows of $\Theta_{t}^{i}$, in view of Definition \ref{def:Theta-Phi}, $\widetilde{\Theta}_{t}^{\pi(i)}$ is well-defined with respect to the permutations $\pi$ and $\overbar{\pi}$. After the permutation, the sufficient condition $\widetilde{\Theta}_{t}^{\pi(i)} \mathbf{P}_{t,s}^{\pi(i)} = \widehat{\mathbf{P}}_{t,s}^{\overbar{\pi}(i)} \widetilde{\Theta}_{s}^{\pi(i)}$ still holds. Thus the corresponding multivariate process remains a Markov structure with respect to the permutation of $\lbrace Y^{1}, Y^{2},\ldots ,Y^{m}\rbrace$.
\end{proof}

Next, we prove the first important property.

\begin{theorem}
The systemic dependence measure is invariant with respect to the permutation of the components of $X$.
\end{theorem}

\begin{proof}
In view of Proposition \ref{prop:perSimilar} and Proposition \ref{prop:perSimilarComponent}, we know that a Markov structure is independent of the ordering of the Markov chains within the collection and the state space of the Markov chain member. All that matters is whether the generators are being tracked correspondingly. Thus, we have that
\begin{equation*}
\nu^{z} \left( T,t,h,X^{\mathcal{I}},x\right) = \nu^{z} \left( T,t,h,X_{\pi}^{\mathcal{I}}, \pi( x)\right) 
\end{equation*}
and 
\begin{equation*}
\nu^{z} \left( T,t,h,X^{\mathcal{D}},x\right) = \nu^{z} \left( T,t,h,X_{\pi}^{\mathcal{D}}, \pi( x)\right) .
\end{equation*}
We conclude the proof.
\end{proof}

\begin{corollary}
The systemic instability measure is invariant with respect to the permutation of the components of $X$.
\end{corollary}

\begin{proof}
Since the relative entropy is invariant with respect to the permutation of the components of $X$, the result follows.
\end{proof}

We end this section with the law invariance property.

\begin{theorem}[Law invariance]\label{thm:srmLI}
Let $X^{\mathcal{D}}$ and $X^{\mathcal{D}^{\prime}}$ be Markov structures for $\lbrace Y^{1},Y^{2},\ldots ,Y^{m}\rbrace$. If $X^{\mathcal{D}}$ and $X^{\mathcal{D}^{\prime}}$ have the same law with respect to $\mathbb{P}$, then we have
\begin{equation*}
\rho^{z} \left( T,t,h,X^{\mathcal{D}},x\right) =\rho^{z} \left( T,t,h,X^{\mathcal{D}^{\prime}},x\right),\quad z,x\in\mathcal{K}^{m},\ 0\leq t\leq T,\ h\in\mathcal{M}.
\end{equation*}
\end{theorem}

\begin{proof}
Note that $X^{\mathcal{D}},X^{\mathcal{D}^{\prime}}$ are multivariate Markov chains. We denote by $\Lambda_{u}^{\mathcal{D}}$ and $\Lambda_{u}^{\mathcal{D}^{\prime}},\ u\geq 0$ the infinitesimal generators of $X^{\mathcal{D}}$ and $X^{\mathcal{D}^{\prime}}$, respectively. Markov chains $X^{\mathcal{D}},X^{\mathcal{D}^{\prime}}$ have the same finite-dimensional distribution if and only if $\Lambda_{u}^{\mathcal{D}} =\Lambda_{u}^{\mathcal{D}^{\prime}},\ u\geq 0$. Besides, $\Lambda_{u}^{\mathcal{D}} =\Lambda_{u}^{\mathcal{D}^{\prime}},\ u\geq 0$ if and only if $\mathbf{P}_{t,s}^{X^{\mathcal{D}}}=\mathbf{P}_{t,s}^{X^{\mathcal{D}^{\prime}}},\ 0\leq t\leq s$. Then we have
\begin{equation*}
\Lambda_{u}^{i} = \Theta_{u}^{i} \Lambda_{u}^{\mathcal{D}} \Phi^{i} =\Theta_{u}^{i} \Lambda_{u}^{\mathcal{D}^{\prime}} \Phi^{i},\quad u\geq 0,\ i\in\mathcal{M}.
\end{equation*}
It follows that the corresponding generator of independent structure is identical,
\begin{equation*}
\sum _{j=1}^{m} \mathrm{I} _{\Lambda_{u}^{j}}^{\otimes m,j} =\sum _{j=1}^{m} \mathrm{I}\otimes \cdots \otimes \Lambda_{u}^{j} \otimes \cdots \otimes \mathrm{I},\quad u\geq 0.
\end{equation*}
For any $z\in\mathcal{K},\ 0\leq t\leq T,\ h\in\mathcal{M}$, we have
\begin{align*}
&\rho^{z} \left( T,t,h,X^{\mathcal{D}},x\right) \\
&=\mathbb{P} \left( \sum _{j=1}^{m} \mathbbm{1}_{\left\lbrace X_{T}^{\mathcal{D},j}=z_{j}\right\rbrace} \geq h\ \bigg\rvert\ X_{t}^{\mathcal{D}}=x \right)
 -\mathbb{P} \left( \sum _{j=1}^{m} \mathbbm{1}_{\left\lbrace X_{T}^{\mathcal{I},j}=z_{j}\right\rbrace} \geq h\ \bigg\rvert\ X_{t}^{\mathcal{I}}=x \right) \\
 &= \mathbb{P} \left( \sum _{j=1}^{m} \mathbbm{1}_{\left\lbrace X_{T}^{\mathcal{D}^{\prime},j}=z_{j}\right\rbrace} \geq h\ \bigg\rvert\ X_{t}^{\mathcal{D}^{\prime}}=x \right)
 -\mathbb{P} \left( \sum _{j=1}^{m} \mathbbm{1}_{\left\lbrace X_{T}^{\mathcal{I},j}=z_{j}\right\rbrace} \geq h\ \bigg\rvert\ X_{t}^{\mathcal{I}}=x \right) \\
 &= \rho^{z} \left( T,t,h,X^{\mathcal{D}^{\prime}},x\right),
\end{align*}
where the second equality comes from the fact that the semigroups of $X^{\mathcal{D}},X^{\mathcal{D}^{\prime}}$ are the same.
\end{proof}

\begin{corollary}
The systemic instability measure is law-invariant.
\end{corollary}

\begin{proof}
The second term in \eqref{eq:scmRealized} is law-invariant due to $\mathbf{P}_{t,s}^{X^{\mathcal{D}}}=\mathbf{P}_{t,s}^{X^{\mathcal{D}^{\prime}}},\ 0\leq t\leq s$. Together with Theorem \ref{thm:srmLI} we conclude the proof. 
\end{proof}

\section{Numerical results}\label{sec:numerical}
In the case of Examples in \cite{ysc17phd} and \cite{bjn13intricacies}, we have full knowledge about how the initial distribution of a bivariate Markov chain and the algebraic structures of infinitesimal generator $\left( \Lambda_{u},u\geq 0\right)$ of $X$ determine different types of Markovian consistency. Therefore, we will follow up on those examples here. We present in this section the numerical study on the systemic instability measure of various Markov structures for a collection of Markov chains $\lbrace Y^{1},Y^{2}\rbrace$. The main objective is to explain financial meanings of various dependence structures and to study the robustness of proposed measures in the simple setting.

\subsection{Model specification}

Here, we consider a financial system of two financial institutions. We denote by $Y^{i}$ the credit rating process of the $i$th financial institution, $i=1,2$. Instead of taking all credit ratings as the state space of $Y^{i}$, we presume several credit events as default. Thus the credit status of each financial institution is in either non-default or default. State $0$ represents the non-default state, and state $1$ stands for the default state. Moreover, for simplicity, we suppose that once the financial institution defaults, it cannot return to the non-default state. It means that state $1$ is an absorbing state for every $Y^{i}$. We further assume that at time $0$ each $Y^{i}$ starts from the non-default state with probability $1$, $\mathbb{P}\left( Y_{0}^{i}=0\right) =1,\ i=1,2$.

Next, let $0\leq v_{1}\leq v_{2}\leq v_{3}\leq \cdots \leq v_{n}<\infty$. We consider time intervals: $[0,v_{1})$, $[v_{1},v_{2})$, $[v_{2},v_{3}),\ldots$, and $[v_{n},\infty )$. Assume that $X^{\mathcal{D}}$ is a bivariate Markov chain with the initial distribution $\mathbb{P}\left( X_{0}=(0,0)\right) =1$ and the infinitesimal generator $\left( \Lambda_{u}^{\mathcal{D}},u\geq 0\right)$,
\begin{align}
\Lambda_{u}^{\mathcal{D}} &=\begingroup
\renewcommand*{\arraystretch}{1.2}%
\renewcommand{\kbldelim}{(}
\renewcommand{\kbrdelim}{)}
\kbordermatrix{
           & (0,0)    & (0,1)   & (1,0)   &(1,1)\cr
  (0,0)  & -(a_{u}+b_{u}+c_{u}) & a_{u} & b_{u} & c_{u} \cr
  (0,1)  & 0 & -d_{u} & 0 & d_{u} \cr
  (1,0)  & 0 & 0 & -f_{u} & f_{u} \cr
  (1,1)  & 0 & 0 & 0 &  0\cr
},\endgroup \label{eq:srmGen}
\end{align}
where for any $0\leq u<\infty$, $a_{u},b_{u},c_{u}, d_{u},f_{u}\geq 0$, and $a_{u},b_{u},c_{u},d_{u},f_{u}$ are piecewise constant on time intervals $[0,v_{1}),\ldots,\ [v_{n},\infty )$. 


Note that, the procedure to construct Markov structures is to take the generator $\left( \Lambda_{u}^{i},u\geq 0\right)$ of $Y^{i},\ i=1,2$, as initial inputs. Then one constructs Markov structures corresponding to $\lbrace Y^{1},Y^{2}\rbrace$. 
If, additionally, we have $a_{u},b_{u},c_{u}>0$, and
\begin{align*}
a_{u} +c_{u} \neq f_{u}   && b_{u} +c_{u} \neq d_{u} ,
\end{align*}
then $X^{\mathcal{D}}$ is weak-only Markovian consistent. Each $X^{\mathcal{D},i}$ is therefore has generator
\begin{equation}
\Lambda_{u}^{i} = \Theta_{u}^{i} \Lambda_{u}^{\mathcal{D}} \Phi^{i},\quad u\geq 0,\ i=1,2.
\end{equation}
Subsequently, we construct the independent infinitesimal generator $\left( \Lambda_{u}^{\mathcal{I}}, u\geq 0\right)$ of $X^{\mathcal{I}}$ by
\begin{equation*}
\Lambda_{u}^{\mathcal{I}} =\sum _{j=1}^{2} \mathrm{I} _{\Lambda_{u}^{j}}^{\otimes 2,j} =\Lambda_{u}^{1} \otimes \mathrm{I} + \mathrm{I} \otimes \Lambda_{u}^{2} .
\end{equation*}
Then we compute the corresponding systemic measures of the Markov structures $X^{\mathcal{D}}$ for $\lbrace Y^{1},Y^{2}\rbrace$.

Throughout the examples, in view of the systemic instability measure defined by \eqref{eq:scmRealized}, we let $t$ vary from $0$ to a finite time horizon $30$ with step size $\Delta t=0.2 \left( \approx \frac{1}{52}\right)$, and choose the monitor window $\Delta T := T-t$. We are interested in the case of $z=(1,1)$ where both components are in default. Also, we are interested in the event when both financial institutions default, $h=2$. Now, the systemic instability measure is a function of the current time $t$ and $X^{\mathcal{D}}$. Thus, the systemic instability measure is given by
\begin{equation*}
\kappa^{(1,1)} \left( t+\Delta T,t,2,X^{\mathcal{D}}, (0,0)\right) =\rho^{(1,1)} \left( t+\Delta T,t,2,X^{\mathcal{D}},(0,0)\right) \times \left( \sum_{y\in\mathcal{K}^{m}} \mathbb{P}\left( X_{t}^{\mathcal{D}} = y\right) \log \frac{\mathbb{P}\left( X_{t}^{\mathcal{D}} = y\right)}{\mathbb{P}\left( X_{t}^{\mathcal{I}} = y\right)}\right),
\end{equation*}
where
\begin{multline}
\rho^{(1,1)} \left( t+\Delta T,t,2,X^{\mathcal{D}},(0,0)\right) \\
 = \mathbb{P} \left( \sum _{j=1}^{2} \mathbbm{1}_{\left\lbrace X_{t+\Delta T}^{\mathcal{D},j}=1\right\rbrace} \geq 2 \ \bigg\rvert\ X_{t}^{\mathcal{D}} =(0,0)\right)
 -\mathbb{P} \left( \sum _{j=1}^{2} \mathbbm{1}_{\left\lbrace X_{t+\Delta T}^{\mathcal{I},j}=1\right\rbrace} \geq 2 \ \bigg\rvert\ X_{t}^{\mathcal{I}} =(0,0)\right) .  \label{eq:srm_t0}
\end{multline}
When the time $t$ proceeds, we have a sequence of values of the systemic instability measure that reflect the financial status of the future time $t+\Delta T$. The specifications of the common model parameters can be found in Table \ref{tbl:parameters}.

\begin{table}[h]
\centering
\caption{Specifications of Model Parameters}\label{tbl:parameters}
\renewcommand{\arraystretch}{1.2}
\begin{tabular}{{c}*{8}{c}c}
\hline \hline
Parameters   & $h$ & $m$ & $\Delta T$ & $v_{1}$ & $v_{2}$ & $v_{3}$ & $v_{4}$ & $v_{5}$  \\
\hline
Value &  2 & 2 & 3 & 6 & 10 & 20 & 26 & 30  \\
\hline
\end{tabular}
\end{table}

\subsection{Examples}

In the following examples, we will mainly analyze the properties of the corresponding systemic instability measure computed based on different dependence structures $\Lambda_{u}^{\mathcal{D}}$. Whereas we will present the composite terms of systemic instability measure in detail for Example \ref{ex:ComFac}.

\begin{example}[Contagious common jumps]\label{ex:ComFac}
Let $Y^{1},Y^{2}$ be Markov chains with generators
\begin{align}\label{eq:marginalEx1}
\Lambda_{u}^{i} &=
\begingroup
\renewcommand*{\arraystretch}{1.2}%
\renewcommand{\kbldelim}{(}
\renewcommand{\kbrdelim}{)}
\kbordermatrix{
           & 0    & 1\cr
  0  & -\lambda_{u}^{i} & \lambda_{u}^{i} \cr
  1  & 0 & 0 \cr
},\endgroup \quad u\geq 0,\ i=1,2,
\end{align}
respectively, where
\begin{align}\label{eq:srmEx1-1}
\begin{split}
\lambda_{u}^{1} &= \frac{c_{u}\left( a+b+c_{u}\right)\mathrm{e}^{-\int_{0}^{u}\left( a+b+c_{v}\right) \d v} +ab\mathrm{e}^{-bu}}{a\mathrm{e}^{-bu} +c_{u}\mathrm{e}^{-\int_{0}^{u}\left( a+b+c_{v}\right) \d v}} \\
\lambda_{u}^{2} &= \frac{c_{u}\left( a+b+c_{u}\right)\mathrm{e}^{-\int_{0}^{u}\left( a+b+c_{v}\right) \d v} +ab\mathrm{e}^{-au}}{b\mathrm{e}^{-au} +c_{u}\mathrm{e}^{-\int_{0}^{u}\left( a+b+c_{v}\right) \d v}} ,
\end{split}
\end{align}
and $a,b,c_{u}>0$, and $c_{u}$ is piecewise constant.

We construct an independence generator by $\Lambda_{u}^{1}$ and $\Lambda_{u}^{2}$,
\begin{align}
\Lambda_{u}^{\mathcal{I}} &= \Lambda_{u}^{1}\otimes \mathrm{I} + \mathrm{I} \otimes \Lambda_{u}^{2} =
\begingroup
\renewcommand*{\arraystretch}{1.2}%
\renewcommand{\kbldelim}{(}
\renewcommand{\kbrdelim}{)}
\kbordermatrix{
           & (0,0)    & (0,1)   & (1,0)   &(1,1)\cr
  (0,0)  & -(\lambda_{u}^{1}+\lambda_{u}^{2}) & \lambda_{u}^{2} & \lambda_{u}^{1} & 0 \cr
  (0,1)  & 0 & -\lambda_{u}^{1} & 0 & \lambda_{u}^{1} \cr
  (1,0)  & 0 & 0 & -\lambda_{u}^{2} & \lambda_{u}^{2} \cr
  (1,1)  & 0 & 0 & 0 &  0\cr
}.\endgroup \label{eq:srmEx1-ind}
\end{align}
The generator $\Lambda_{u}^{\mathcal{I}}$ satisfies condition \ref{cnd:m}. Now, we solve for a solution $\left( \Lambda_{u}^{\mathcal{D}}, u\geq 0\right)$ of
\begin{equation*}
\Lambda_{u}^{i} = \Theta_{u}^{i} \Lambda_{u}^{\mathcal{D}} \Phi^{i},\quad u\geq 0,\ i=1,2,
\end{equation*}
where $\left( \Lambda_{u}^{\mathcal{D}}, u\geq 0\right)$ is of the form
\begin{align}
\Lambda_{u}^{\mathcal{D}} &= \begingroup
\renewcommand*{\arraystretch}{1.2}%
\renewcommand{\kbldelim}{(}
\renewcommand{\kbrdelim}{)}
\kbordermatrix{
           & (0,0)    & (0,1)   & (1,0)   &(1,1)\cr
  (0,0)  & -(a+b+c_{u}) & a & b & c_{u} \cr
  (0,1)  & 0 & -b & 0 & b \cr
  (1,0)  & 0 & 0 & -a & a \cr
  (1,1)  & 0 & 0 & 0 &  0\cr
}.\endgroup \label{eq:srmEx1-dep}
\end{align}
It should be noted that since $c_{u}>0$, a bivariate Markov chain $X^{\mathcal{D}}=\left( X^{\mathcal{D},1},X^{\mathcal{D},2}\right)$ generated by \eqref{eq:srmEx1-dep} is a weak-only Markov structure for $\lbrace Y^{1},Y^{2}\rbrace$. If we take $c_{u}=0$ for all $u\geq 0$ in \eqref{eq:srmEx1-dep}, then we have the independence structure. The parameter $c_{u}$ in \eqref{eq:srmEx1-dep} is the key element that captures the \textit{contagion} of simultaneous jumps between the components. Thus, our first goal is to study how the parameter $c_{u}$ affects the systemic instability measure. 

Next, we consider another generator $\Lambda_{u}^{\mathcal{S}}$ of the form,
\begin{align}
\Lambda_{u}^{\mathcal{S}} &= \begingroup
\renewcommand*{\arraystretch}{1.2}%
\renewcommand{\kbldelim}{(}
\renewcommand{\kbrdelim}{)}
\kbordermatrix{
           & (0,0)    & (0,1)   & (1,0)   &(1,1)\cr
  (0,0)  & -\left( \lambda_{u}^{1}+\lambda_{u}^{2}-g_{u}\right) & \lambda_{u}^{2}-g_{u} & \lambda_{u}^{1}-g_{u} & g_{u} \cr
 (0,1)  & 0 & -\lambda_{u}^{1} & 0 & \lambda_{u}^{1} \cr
  (1,0)  & 0 & 0 & -\lambda_{u}^{2} & \lambda_{u}^{2} \cr
  (1,1)  & 0 & 0 & 0 &  0\cr
},\endgroup \quad u\geq 0,\label{eq:srmEx1-strM}
\end{align}
where $0\leq g_{u}\leq \min \left( \lambda_{u}^{1},\lambda_{u}^{2}\right)$. The generator $\Lambda_{u}^{\mathcal{S}}$ satisfies condition \ref{cnd:m} as well, and a Markov chain $X^{\mathcal{S}}$ generated by $\left( \Lambda_{u}^{\mathcal{S}},u\geq 0\right)$ is a strong Markov structure for $\lbrace Y^{1},Y^{2}\rbrace$. Moreover, if $g_{u}=0$, \eqref{eq:srmEx1-strM} reduces to independence structure \eqref{eq:srmEx1-ind}. Note that, by construction, the $i$th component of a bivariate chain generated respectively by either $\Lambda_{u}^{\mathcal{D}},\Lambda_{u}^{\mathcal{I}}$ or $\Lambda_{u}^{\mathcal{S}}$ has identical prescribed law $\left( \Lambda_{u}^{i}, u\geq 0\right)$ in \eqref{eq:marginalEx1}.

In the first part of the numerical analysis, we restrict ourselves to the structure \eqref{eq:srmEx1-dep} and show the compositions of systemic instability measure. For simplicity, we assume that the financial system has finite lifetime $T=30$ and let $t$ vary. Two scenarios are considered: piecewise increasing $c_{u}$ and piecewise decreasing $c_{u}$. The parameters are given in Table \ref{tbl:parEx1}.

\begin{table}[h]
\centering
\caption{Parameters for Example \ref{ex:ComFac} (Contagious Common Jumps): $\kappa^{(1,1)} \left( 30,t,2,X^{\mathcal{D}}, (0,0)\right)$}\label{tbl:parEx1}
\renewcommand{\arraystretch}{1.2}
\begin{tabular}{{c}*{7}{c}c}
\hline \hline
\multirow{2}{*}{Parameters} & \multicolumn{5}{c}{Time Periods} \\
\cline{2-6}
& $[0,3)$ & $[3,10)$  & $[10,30)$ & $[30,\infty)$  &  \\
\hline
$a$ & 0.01 & 0.01 & 0.01 & 0.01 &    \\
$b$ & 0.02 & 0.02 & 0.02 & 0.02 &    \\
Scenario 1: $c_{u}$ piecewise increasing & 0.08 & 0.15 & 0.2 & 0.2 &    \\
Scenario 2: $c_{u}$ piecewise decreasing & 0.08 & 0.03 & 0.003 & 0.003 &    \\
\hline
\end{tabular}
\end{table}


Figure \ref{fig:scmEx1-1} on page \pageref{fig:scmEx1-2} shows the result of $\kappa^{(1,1)} \left( 30,t,2,X^{\mathcal{D}}, (0,0)\right)$. We observe that at the initial time $t=0$ and the termination $t=30$, the values of the systemic instability measure are both zero. They come from different reasons. At time $0$, it results from the same initial distribution. However, at time $T$, it is caused by the end of the simulated processes. Note that we fix parameter $T$ in this simulation. Although in both cases (piecewise increasing and piecewise decreasing), we have the same parameters in the first period $[0,3)$, in general, we should not expect the same level of systemic instability measure, that depends on the entire simulated processes.

\begin{figure}
\begin{subfigure}[h]{0.49\linewidth}
\includegraphics[width=3.5in]{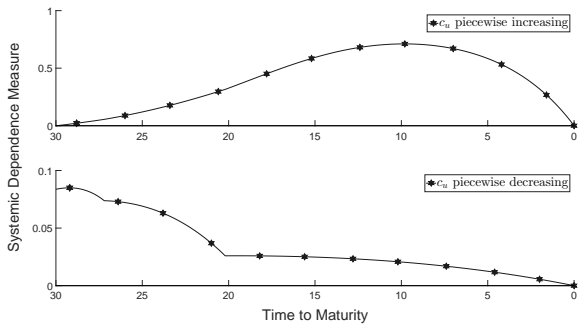}
\caption{Systemic dependence measure}
\end{subfigure}
\hfill
\begin{subfigure}[h]{0.49\linewidth}
\includegraphics[width=3.5in]{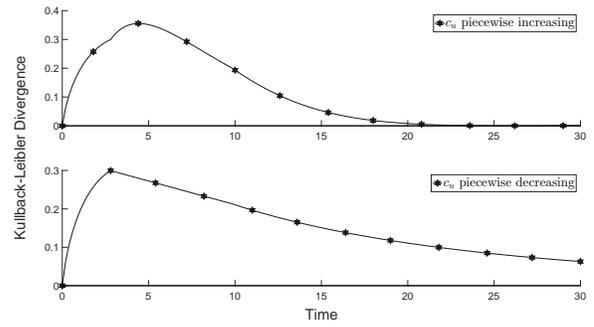}
\caption{Kullback-Leibler divergence}
\end{subfigure}
\hfill
\begin{subfigure}[h]{1\linewidth}
\centering
\includegraphics[width=3.5in]{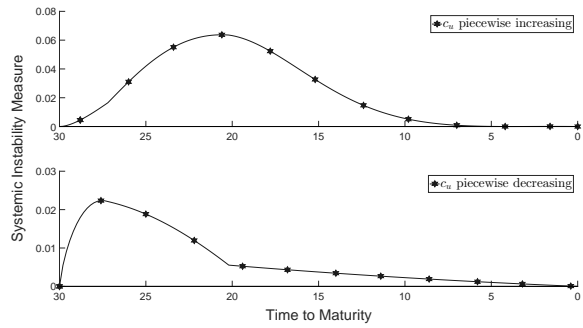}
\caption{Systemic instability measure}
\end{subfigure}

\caption{Systemic Dependence Measure, Kullback-Leibler Divergence, and Systemic Intability Measure for Example \ref{ex:ComFac} (Contagious Common Jumps)}\label{fig:scmEx1-1}
\end{figure}

In the second part of the numerical analysis, we examine the systemic instability measure for two different dependence structures \eqref{eq:srmEx1-dep} and \eqref{eq:srmEx1-strM}. In addition to the parameters in Table \ref{tbl:parEx1}, we take parameter $g_{u} =\eta \min \left( \lambda_{u}^{1},\lambda_{u}^{2}\right)$ with $\eta =0, 0.5, 0.8, 1$. If $\eta=0$, the structure \eqref{eq:srmEx1-dep} becomes the independence structure \eqref{eq:srmEx1-ind}. 

\begin{figure}[!htb]
\centering
\includegraphics[width=5in]{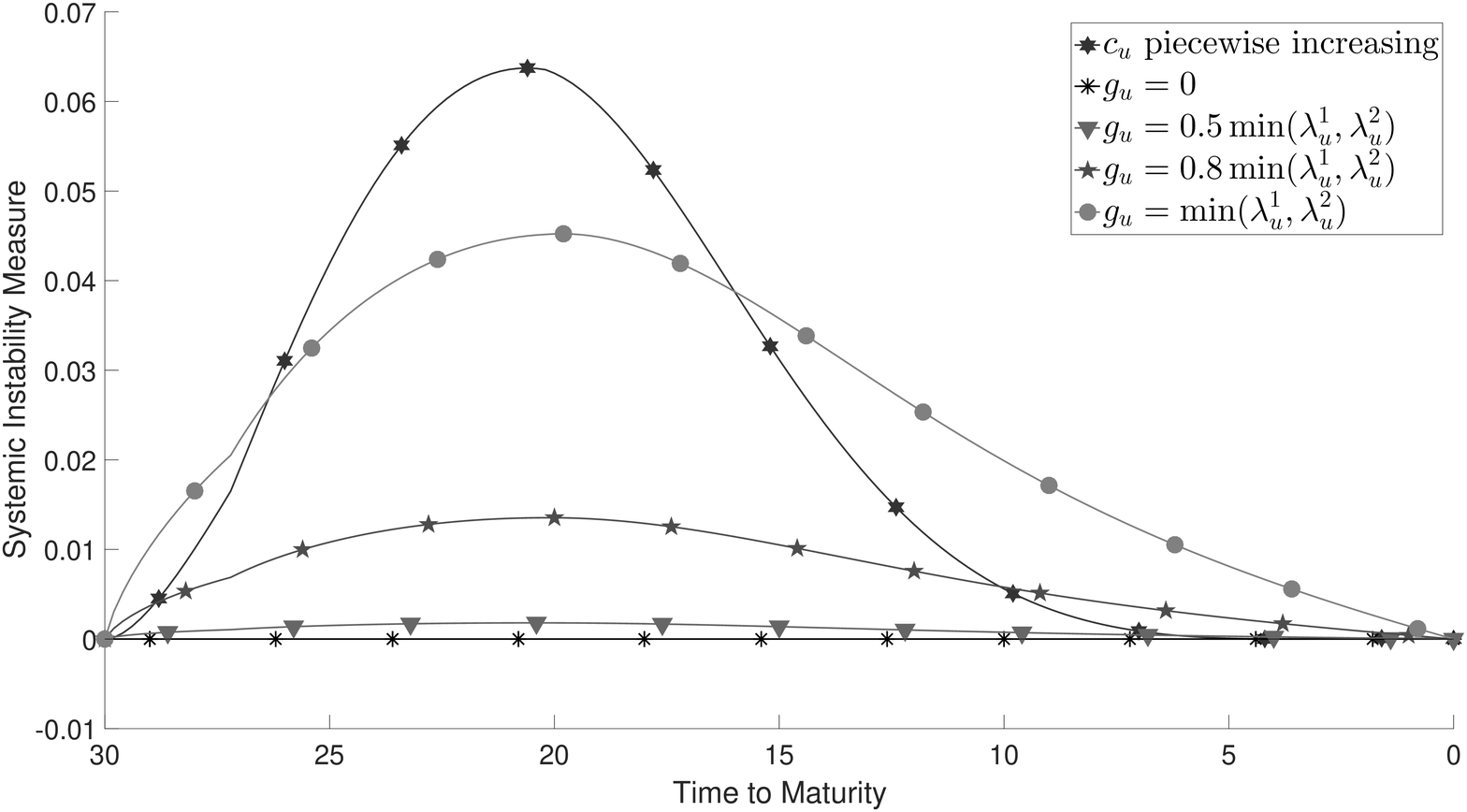}
\caption{Systemic Instability Measure for Example \ref{ex:ComFac} (Contagious Common Jumps) with Piecewise Increasing $c_{u}$}\label{fig:scmEx1-2}
\end{figure}

\begin{figure}[!htb]
\centering
\includegraphics[width=5in]{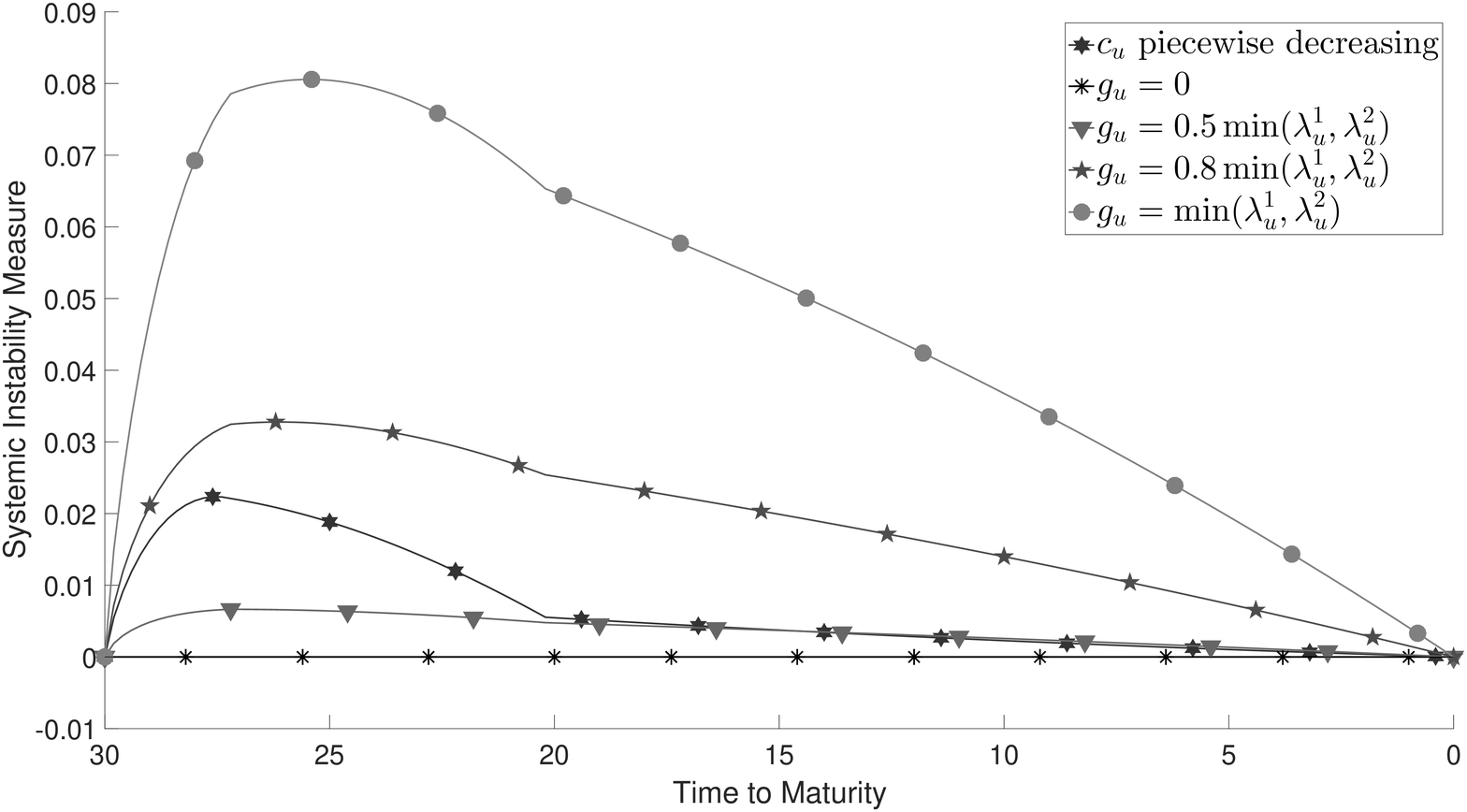}
\caption{Systemic Instability Measure for Example \ref{ex:ComFac} (Contagious Common Jumps) with Piecewise Decreasing $c_{u}$}\label{fig:scmEx1-3}
\end{figure}

In Figure \ref{fig:scmEx1-2} on page \pageref{fig:scmEx1-2} and Figure \ref{fig:scmEx1-3} on page \pageref{fig:scmEx1-3}, we compare the systemic instability  measure of the weak-only Markov structure to the strong Markov structures. It should be noted that when $\eta=1$, one of the components cannot jump individually. For instance, if $\min \left( \lambda_{u}^{1},\lambda_{u}^{2}\right) =\lambda_{u}^{2}$, then the probability for $X^{\mathcal{S},2}$ jumping individually to state $1$ is zero. In such a case, the level of systemic instability measure is highest. We argue that, since condition \ref{cnd:m} is satisfied, the state of $X^{\mathcal{S},1}$ will not have an influence on how $X^{\mathcal{S},2}$ changes its state. However, if $X^{\mathcal{S},2}$ changes to state $1$, $X^{\mathcal{S},1}$ must jump to state $1$ as well. Financially speaking, the corresponding financial institution $X^{\mathcal{S},2}$ will not default individually. However, if $X^{\mathcal{S},2}$ defaults, then $X^{\mathcal{S},1}$ must default as well. The component $X^{\mathcal{S},2}$ contributes significantly to systemic risk. 

After visualizing the basic behaviors of systemic instability measure, let us switch gears to more practical case: $\kappa^{(1,1)} \left( t+\Delta T,t,2,X^{\mathcal{D}}, (0,0)\right)$. As $\Delta T$ is fixed, the measure evaluates a fixed monitor period, from time $t$ to time $t+\Delta T$, of the systemic condition at time $t$. It will allow us to routinely monitor the future condition of the financial system, for instance, on the weekly basis. We consider two scenarios for the parameter $c_{u}$ fluctuating in different patterns. In the first scenario, $c_{u}$ starts with small value, then it becomes large and go down. The interpretation is that both financial institutions begins with good conditions. Then they deteriorate simultaneously through the common jumps factor, and become better again through the common jumps factor. Instead, the second scenario, $c_{u}$ will start with large value and cycle. We collect the parameters in Table \ref{tbl:parEx1-2}.

\begin{table}[h]
\centering
\caption{Parameters for Example \ref{ex:ComFac} (Contagious Common Jumps): $\kappa^{(1,1)} \left( t+\Delta T,t,2,X^{\mathcal{D}}, (0,0)\right)$}\label{tbl:parEx1-2}
\renewcommand{\arraystretch}{1.2}
\begin{tabular}{{c}*{7}{c}c}
\hline \hline
\multirow{2}{*}{Parameters} & \multicolumn{7}{c}{Time Periods} \\
\cline{2-8}
& $[0,6)$ & $[6,10)$  & $[10,20)$ & $[20,26)$ & $[26,30)$ & $[30,\infty)$  &  \\
\hline
$a$ & 0.01 & 0.01 & 0.01 & 0.01 & 0.01 & 0.01 &    \\
$b$ & 0.02 & 0.02 & 0.02 & 0.02 & 0.02 & 0.02 &    \\
\multicolumn{1}{l}{Scenario 1: $c_{u}$} & 0.01 & 0.09 & 0.03 & 0.12 & 0.04 & 0.04 \\
\multicolumn{1}{l}{Scenario 2: $c_{u}$}     & 0.12 & 0.09 & 0.03 & 0.09 & 0.05 & 0.05  \\
\hline
\end{tabular}
\end{table}

In Figure \ref{fig:scmEx1-4} on page \pageref{fig:scmEx1-4}, all sequences of systemic instability measure fluctuate. As the monitor window $\Delta T=3$, the systemic instability measure uproars at $t=3$ reflecting the sudden change of $c_{u}$ in period $[6,10)$. Analogously, the measure decreases at $t=7$ indicating the jump of $c_{u}$ in period $[10,20)$.  Similar results can be observed in Figure \ref{fig:scmEx1-5} on page \pageref{fig:scmEx1-5}. We would like to emphasize that one Markov structure may behave more stable or unstable than other Markov structures throughout the entire tracking period. 

\begin{figure}[!htb]
\centering
\includegraphics[width=5in]{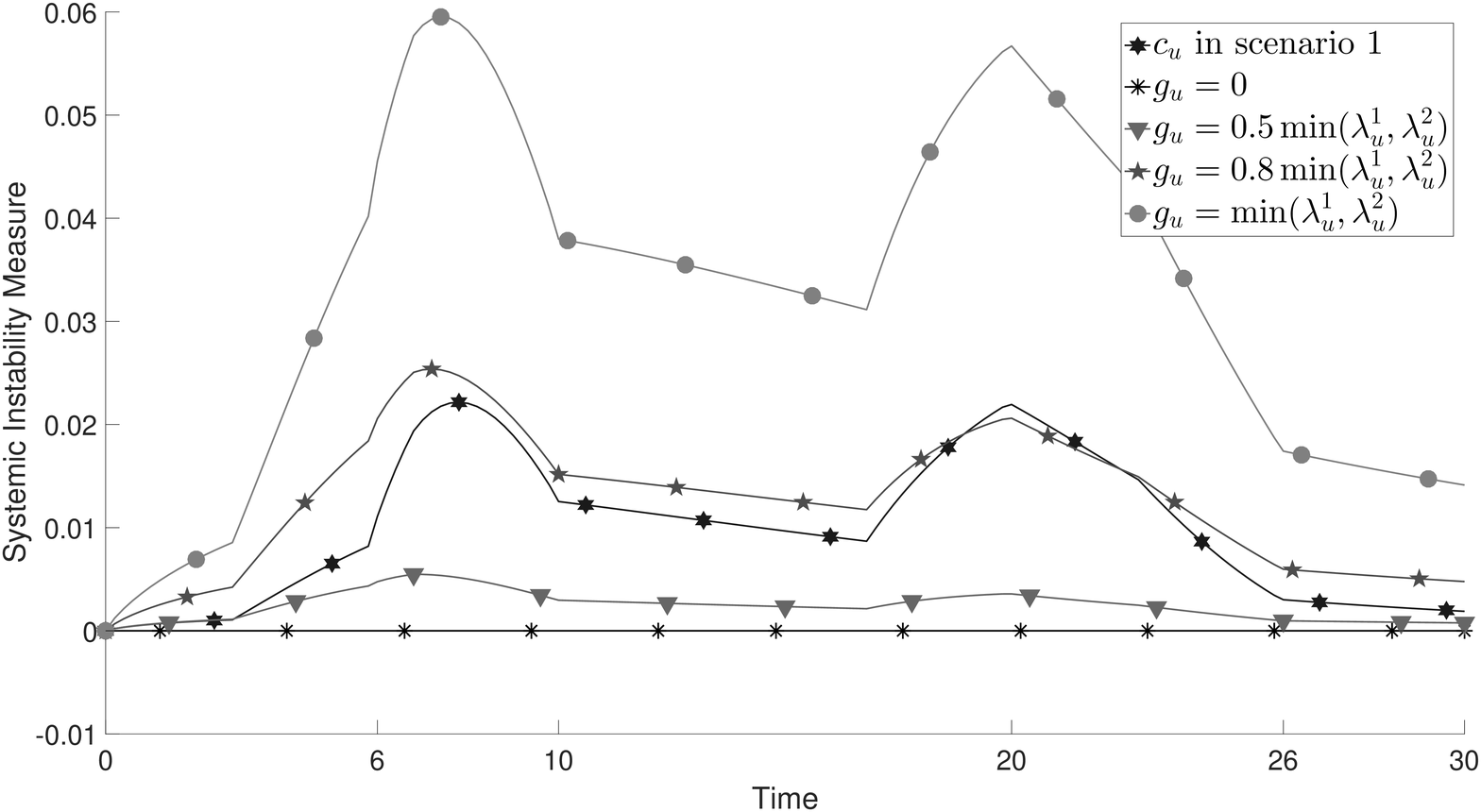}
\caption{Systemic Instability Measure for Example \ref{ex:ComFac} (Contagious Common Jumps): Scenario 1}\label{fig:scmEx1-4}
\end{figure}

\begin{figure}[!htb]
\centering
\includegraphics[width=5in]{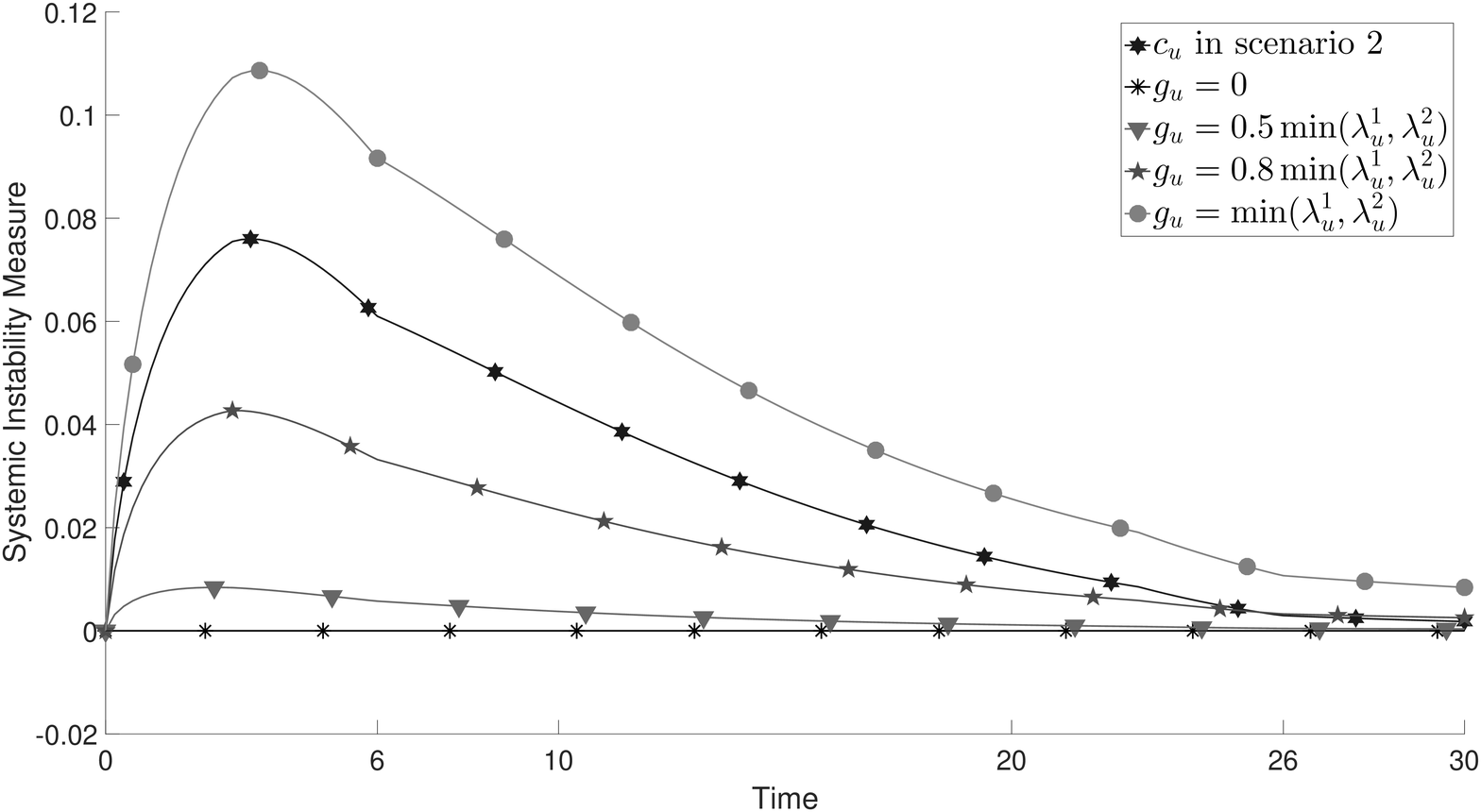}
\caption{Systemic Instability Measure for Example \ref{ex:ComFac} (Contagious Common Jumps): Sceneario 2}\label{fig:scmEx1-5}
\end{figure}

Now, we use the same data in Table \ref{tbl:parEx1-2} while changing the lengths of monitor window $\Delta T$, $\Delta
 T = 0.6, 1, 3, 5$. The comparison is presented in Figure \ref{fig:scmEx1-6} on page \pageref{fig:scmEx1-6}. The sequence of systemic instability measure with shorter monitor window $\Delta T$ is embraced by the longer monitor window $\Delta T$,
\begin{equation*}
\abs{\kappa^{(1,1)} \left( t+\Delta T_{1},t,2,X^{\mathcal{D}}, (0,0)\right)} \leq \abs{\kappa^{(1,1)} \left( t+\Delta T_{2},t,2,X^{\mathcal{D}}, (0,0)\right)},\quad 0\leq \Delta T_{1}\leq \Delta T_{2} <\infty.
\end{equation*} 
From the perspective of risk management, this property should be expected as the uncertainty of a financial system for a longer period is generally higher.

\begin{figure}
\begin{subfigure}[h]{0.49\linewidth}
\includegraphics[width=3.5in]{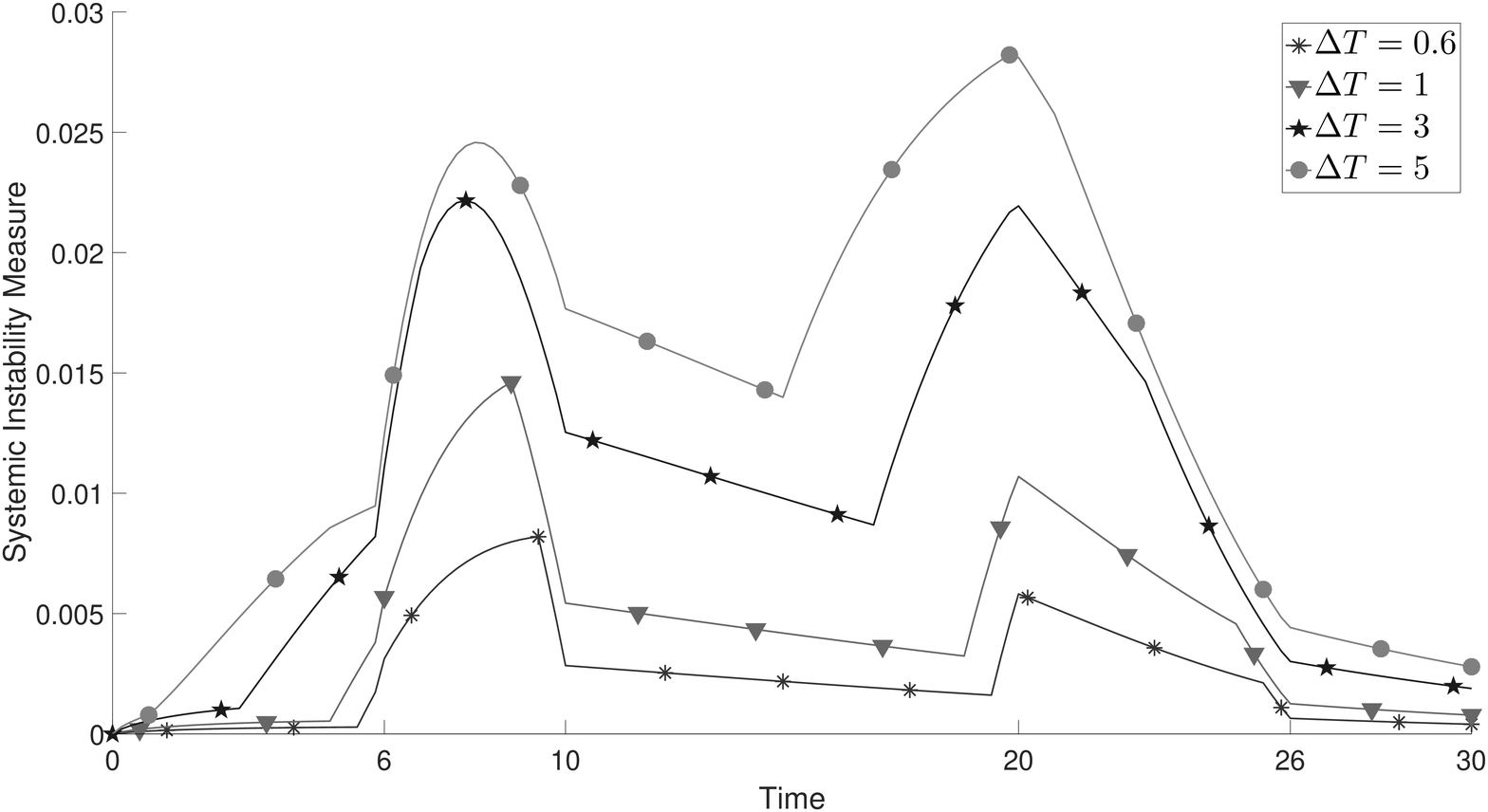}
\caption{Scenario 1}
\end{subfigure}
\hfill
\begin{subfigure}[h]{0.49\linewidth}
\includegraphics[width=3.5in]{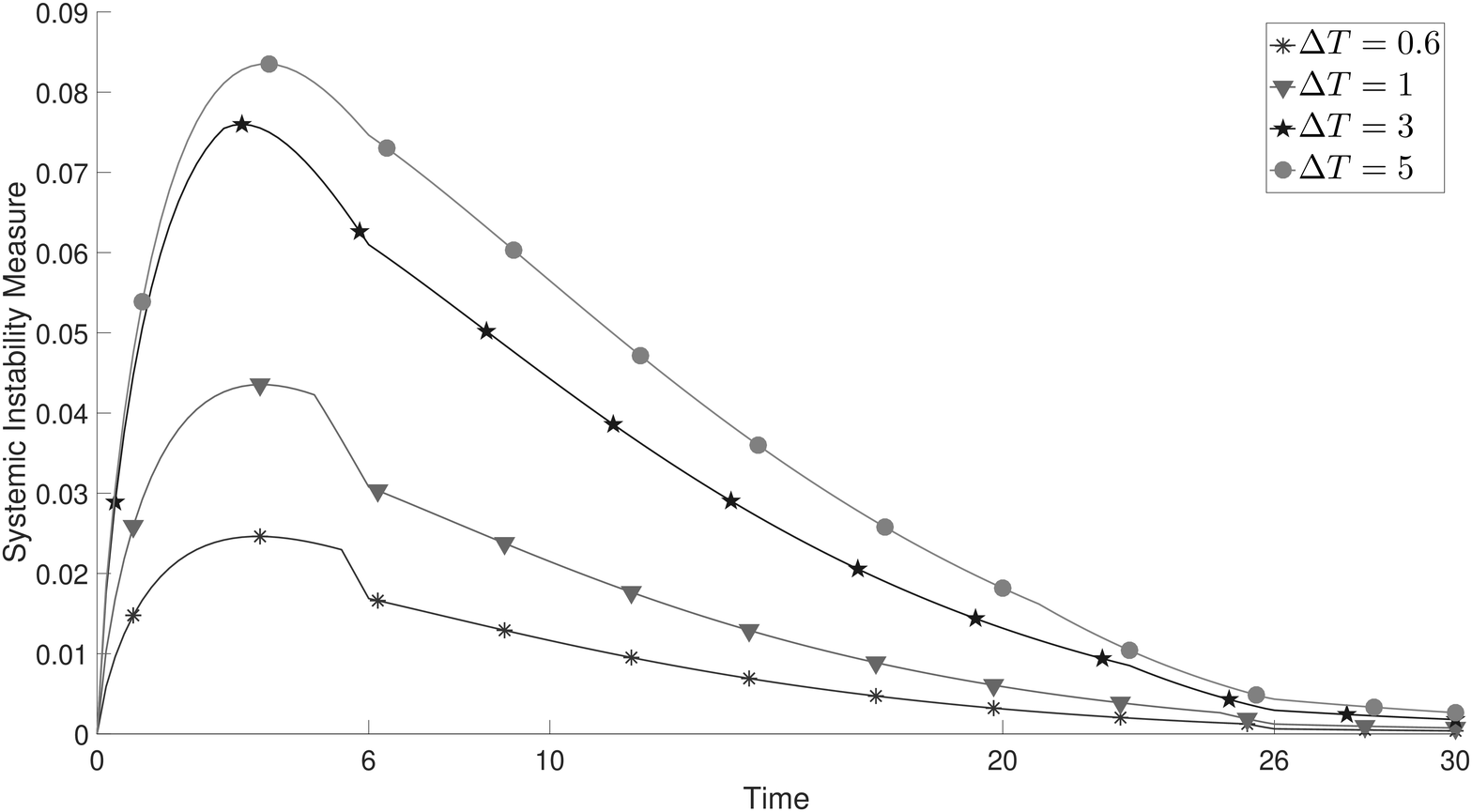}
\caption{Scenario 2}
\end{subfigure}
\centering
\caption{Systemic Instability Measure for Example \ref{ex:ComFac} (Contagious Common Jumps) with Different Lengths of Monitor Window $\Delta T$ }\label{fig:scmEx1-6}
\end{figure}

\end{example}

\begin{example}[Extreme contagion]\label{ex:scmExtrCont}
Assume that $Y^{i}$ is generated by $\Lambda_{u}^{i}$,
\begin{align*}
\Lambda_{u}^{i} &=
\begingroup
\renewcommand*{\arraystretch}{1.2}%
\renewcommand{\kbldelim}{(}
\renewcommand{\kbrdelim}{)}
\kbordermatrix{
           & 0    & 1\cr
  0  & -c_{u} & c_{u} \cr
  1  & 0 & 0 \cr
},\endgroup \quad u\geq 0,\ i=1,2.
\end{align*}
We solve for the solution $\left( \Lambda_{u}^{\mathcal{D}},u\geq 0\right)$ of the form,
\begin{align}
\Lambda_{u}^{\mathcal{D}} &= \begingroup
\renewcommand*{\arraystretch}{1.2}%
\renewcommand{\kbldelim}{(}
\renewcommand{\kbrdelim}{)}
\kbordermatrix{
           & (0,0)    & (0,1)   & (1,0)   &(1,1)\cr
  (0,0)  & -c_{u} & 0 & 0 & c_{u} \cr
  (0,1)  & 0 & 0 & 0 & 0 \cr
  (1,0)  & 0 & 0 & 0 & 0 \cr
  (1,1)  & 0 & 0 & 0 & 0 \cr
},\endgroup \label{eq:srmEx2}
\end{align}
where $c_{u}>0$ and piecewise constant. In view of \eqref{eq:srmEx2}, if the components jump, they must jump simultaneously. Note that in this structure, the transition probability for individual jump is zero. Hence, we call this structure as \textbf{extreme contagion}.

The parameters of this example are summarized in Table \ref{tbl:parEx2}. We investigate the general properties of the systemic instability measure for the structure \eqref{eq:srmEx2}, and then compare the results to the structure \eqref{eq:srmEx1-strM} with different $g_{u}$'s.

\begin{table}[h]
\centering
\caption{Parameters for Example \ref{ex:scmExtrCont} (Extreme Contagion)}\label{tbl:parEx2}
\renewcommand{\arraystretch}{1.2}
\begin{tabular}{{l}*{7}{c}c}
\hline \hline
\multirow{2}{*}{Parameters} & \multicolumn{7}{c}{Time Periods} \\
\cline{2-8}
& $[0,6)$ & $[6,10)$  & $[10,20)$ & $[20,26)$ & $[26,30)$ & $[30,\infty)$  &  \\
\hline
Scenario 1: $c_{u}$  & $0.01$ & $0.1$ & $0.08$ & $0.05$ & $0.03$ & $0.03$    \\
Scenario 2: $c_{u}$  & $0.08$ & $0.03$ & $0.02$ & $0.04$ &  $0.03$ & $0.03$   \\
\hline
\end{tabular}
\end{table}

In Figure \ref{fig:scmEx2-1} on page \pageref{fig:scmEx2-1} and Figure \ref{fig:scmEx2-2} on page \pageref{fig:scmEx2-2}, an interesting observation is that the systemic instability measure of extreme contagion coincides with the measure obtained by $g_{u} =\min \left( \lambda_{u}^{1},\lambda_{u}^{2}\right)$. The mathematical explanations are given below. First, $Y^{1}$ and $Y^{2}$ have the same law. For the structure \eqref{eq:srmEx2}, the chain does not move if it is not in state $(0,0)$. For the structure \eqref{eq:srmEx1-strM}, condition \ref{cnd:p} is satisfied and the transition probability of individual jumps from state $(0,0)$ is $0$. Thus both structures produce the same levels of systemic instability.

\begin{figure}[!htb]
\centering
\includegraphics[width=5in]{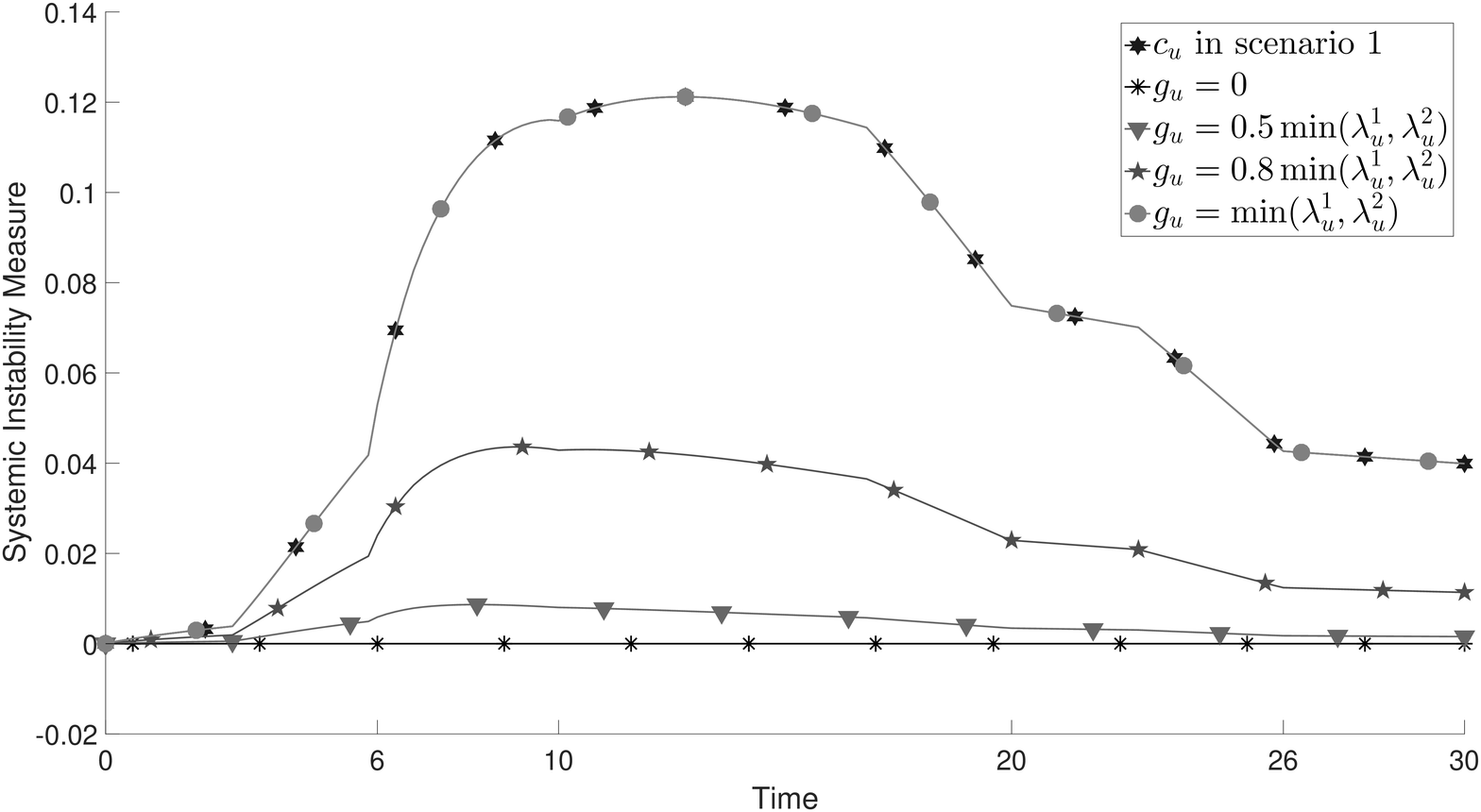}
\caption{Systemic Instability Measure for Example \ref{ex:scmExtrCont} (Extreme Contagion): Scenario 1}\label{fig:scmEx2-1}
\end{figure}

\begin{figure}[!htb]
\centering
\includegraphics[width=5in]{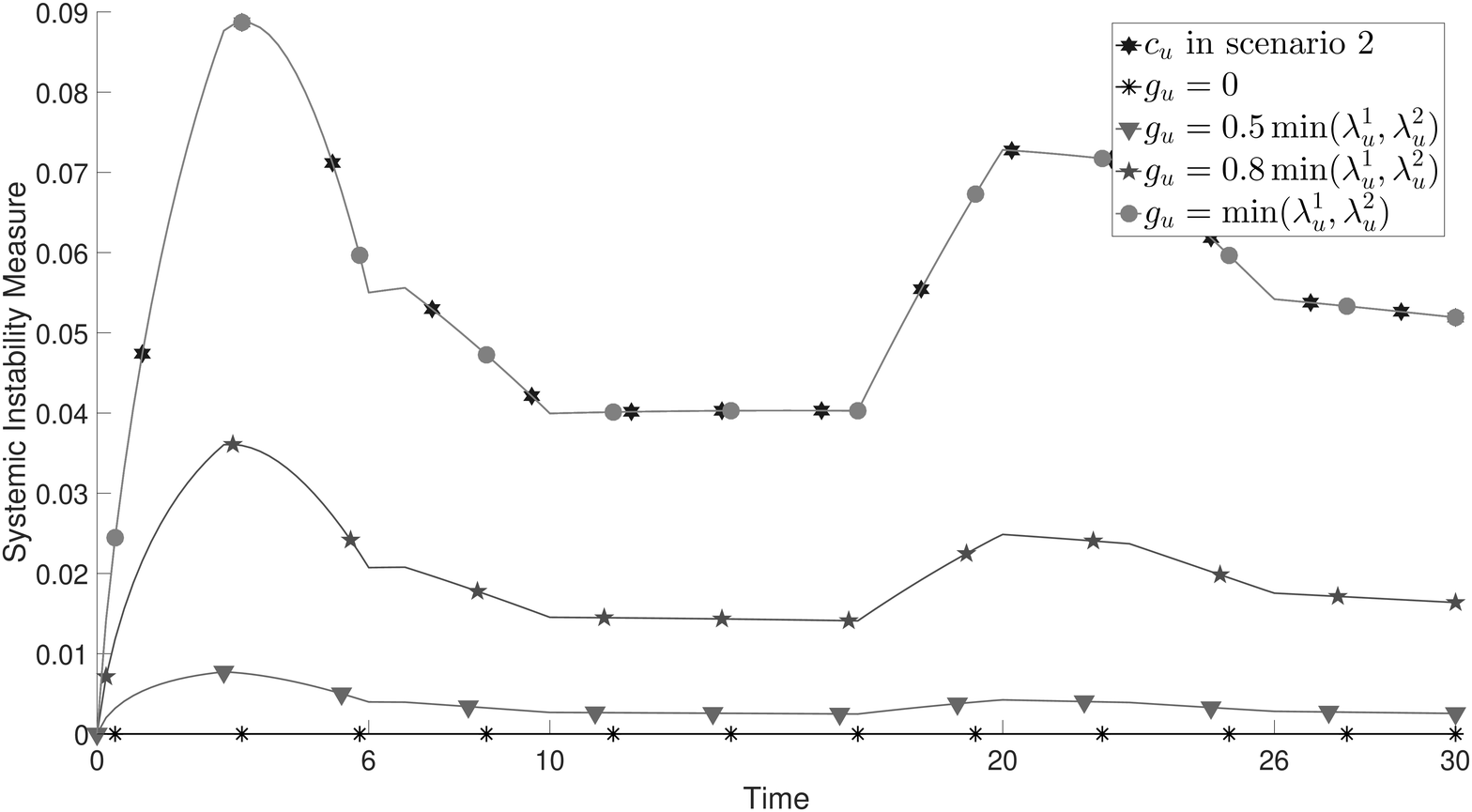}
\caption{Systemic Instability Measure for Example \ref{ex:scmExtrCont} (Extreme Contagion): Scenario 2}\label{fig:scmEx2-2}
\end{figure}

\end{example}

\begin{example}[Extreme anti-contagion]\label{ex:scmExtrAnti}
Consider Markov chains $Y^{1},Y^{2}$ with generators
\begin{align*}
\Lambda_{u}^{i} &=
\begingroup
\renewcommand*{\arraystretch}{1.2}%
\renewcommand{\kbldelim}{(}
\renewcommand{\kbrdelim}{)}
\kbordermatrix{
           & 0    & 1\cr
  0  & -\lambda_{u}^{i} & \lambda_{u}^{i} \cr
  1  & 0 & 0 \cr
},\endgroup \quad u\geq 0,\ i=1,2,
\end{align*}
respectively, where
\begin{align*}
\begin{split}
\lambda_{u}^{1} &= \frac{b_{u}\left( a_{u}+b_{u}\right) \mathrm{e}^{-\int_{0}^{u} \left( a_{v}+b_{v}\right) \d v}}{a_{u} +b_{u} \mathrm{e}^{-\int_{0}^{u} \left( a_{v}+b_{v}\right) \d v}} \\
\lambda_{u}^{2} &= \frac{a_{u}\left( a_{u}+b_{u}\right) \mathrm{e}^{-\int_{0}^{u} \left( a_{v}+b_{v}\right) \d v}}{b_{u} +a_{u} \mathrm{e}^{-\int_{0}^{u} \left( a_{v}+b_{v}\right) \d v}} ,
\end{split}
\end{align*}
and $a_{u},b_{u}>0$ and piecewise constant.

We solve for a Markov structure $X^{\mathcal{D}}$ for $\left\lbrace Y^{1},Y^{2}\right\rbrace$ with generator $\left( \Lambda_{u}^{\mathcal{D}},u\geq 0\right)$ of the form,
\begin{align}
\Lambda_{u}^{\mathcal{D}} &= \begingroup
\renewcommand*{\arraystretch}{1.2}%
\renewcommand{\kbldelim}{(}
\renewcommand{\kbrdelim}{)}
\kbordermatrix{
           & (0,0)    & (0,1)   & (1,0)   &(1,1)\cr
  (0,0)  & -(a_{u}+b_{u}) & a_{u} & b_{u} & 0 \cr
  (0,1)  & 0 & 0 & 0 & 0 \cr
  (1,0)  & 0 & 0 & 0 & 0 \cr
  (1,1)  & 0 & 0 & 0 & 0 \cr
},\endgroup \label{eq:srmEx3}
\end{align}
where $a_{u},b_{u}>0$.

Unlike the structure \eqref{eq:srmEx2} of extreme contagion, in view of the structure \eqref{eq:srmEx3}, every component jumps individually. It is impossible for both components to jump simultaneously,
\begin{equation}
\mathbb{P} \left( \sum _{j=1}^{2} \mathbbm{1}_{\left\lbrace X_{T}^{\mathcal{D},j}=1\right\rbrace} \geq 2 \ \bigg\rvert\ X_{t}^{\mathcal{D}} =(0,0)\right)  =0 ,\quad t\in [0, T].
\end{equation}
We call this structure as \textbf{extreme anti-contagion}. The theoretical value of the systemic dependence measure is non-positive for any $0\leq t<\infty$. The parameters for this example are given in Table \ref{tbl:parEx3}.

\begin{table}[h]
\centering
\caption{Parameters for Example \ref{ex:scmExtrAnti} (Extreme Anti-contagion)}\label{tbl:parEx3}
\renewcommand{\arraystretch}{1.2}
\begin{tabular}{{l}*{7}{c}c}
\hline \hline
\multirow{2}{*}{Parameters} & \multicolumn{7}{c}{Time Periods} \\
\cline{2-8}
& $[0,6)$ & $[6,10)$  & $[10,20)$ & $[20,26)$ & $[26,30)$ & $[30,\infty)$  &  \\
\hline
Scenario 1: $a_{u}=b_{u}$ & $0.01$ & $0.08$ & $0.05$ & $0.03$ & $0.01$ & $0.01$  & \\
Scenario 2: $a_{u}=b_{u}$ & $0.05$ & $0.02$ & $0.03$ & $0.07$ & $0.05$ & $0.05$ &  \\
\hline
\end{tabular}
\end{table}

In Figure \ref{fig:scmEx3-1} on page \pageref{fig:scmEx3-1} and Figure \ref{fig:scmEx3-2} on page \pageref{fig:scmEx3-2}, every curve stands for the systemic instability measure as a function of $t$. We observe that the systemic instability measures are indeed non-positive for the structure \eqref{eq:srmEx3}, and thus the financial system has favorable systemic dependence and is of systemic benefit. Besides, the financial system shares more favorable systemic benefit when the values of $a_{u},b_{u}$ get larger. 


\begin{figure}[!htb]
\centering
\includegraphics[width=5in]{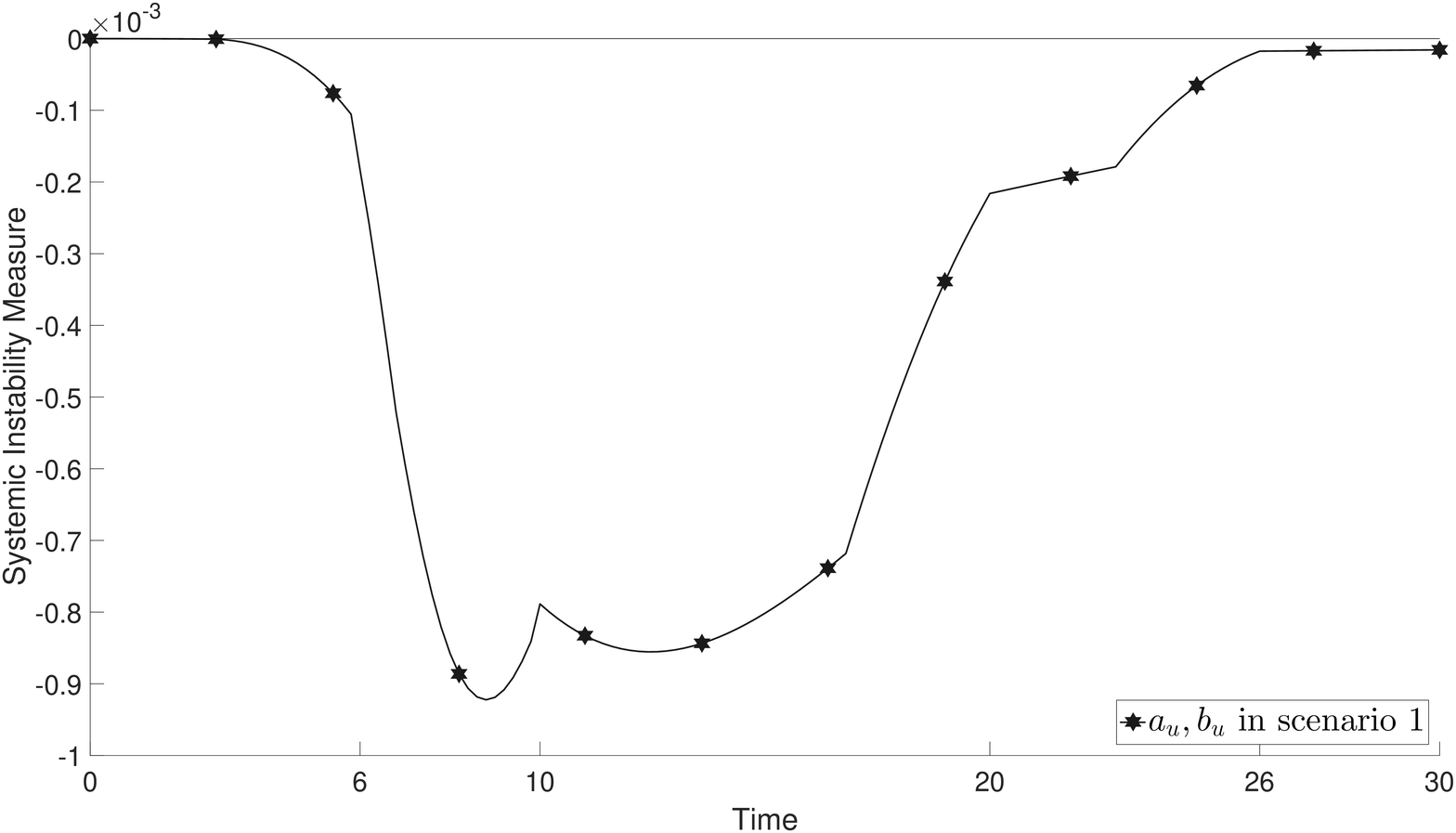}
\caption{Systemic Instability Measure for Example \ref{ex:scmExtrAnti} (Extreme Anti-contagion): Scenario 1}\label{fig:scmEx3-1}
\end{figure}

\begin{figure}[!htb]
\centering
\includegraphics[width=5in]{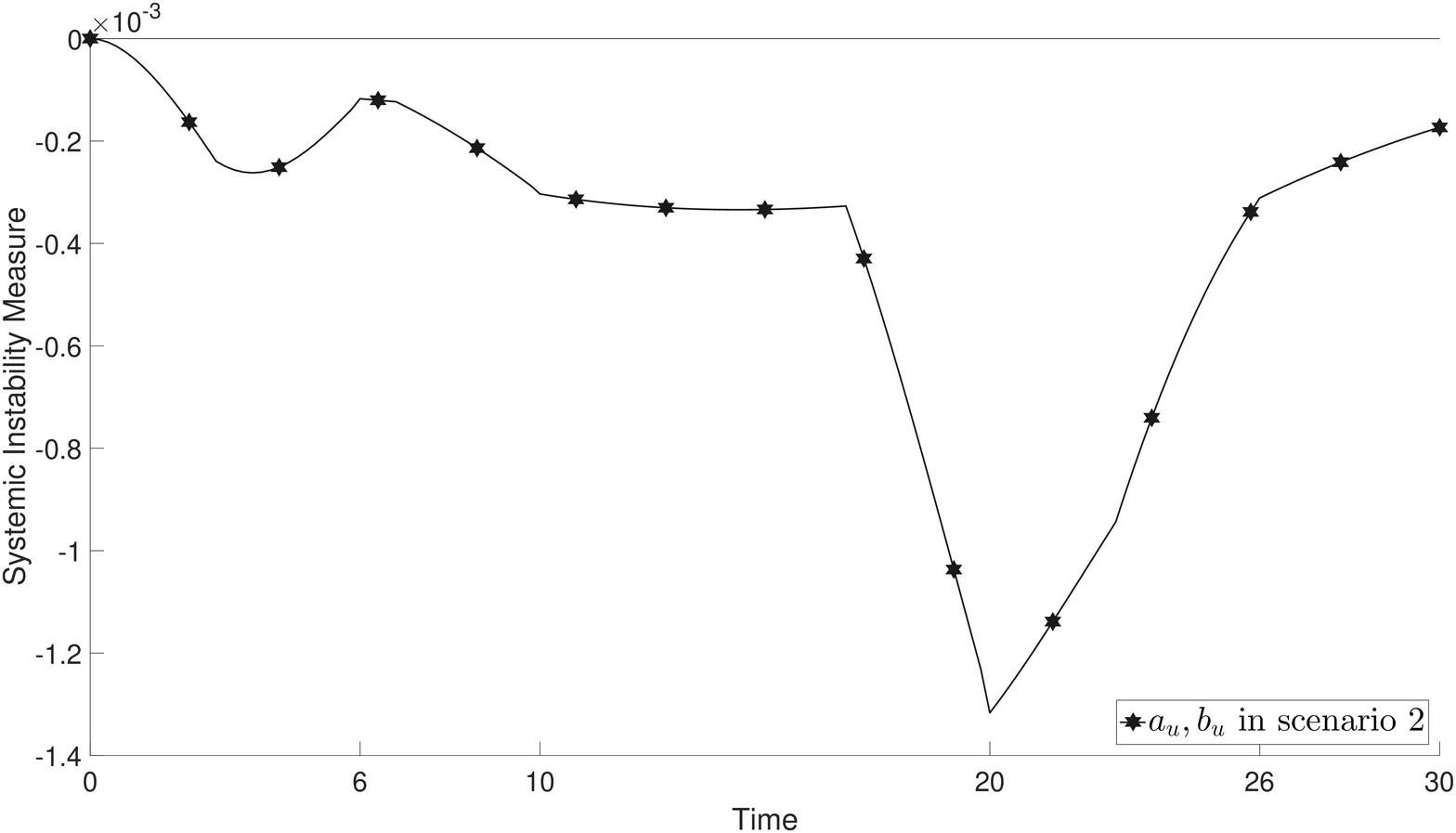}
\caption{Systemic Instability Measure for Example \ref{ex:scmExtrAnti} (Extreme Anti-contagion): Scenario 2}\label{fig:scmEx3-2}
\end{figure}

\end{example}

\begin{example}[Systemic importance]\label{ex:scmSysImp}
Consider Markov chains $Y^{1},Y^{2}$ with generators
\begin{align*}
\Lambda_{u}^{i} &=
\begingroup
\renewcommand*{\arraystretch}{1.2}%
\renewcommand{\kbldelim}{(}
\renewcommand{\kbrdelim}{)}
\kbordermatrix{
           & 0    & 1\cr
  0  & -\lambda_{u}^{i} & \lambda_{u}^{i} \cr
  1  & 0 & 0 \cr
},\endgroup \quad u\geq 0,\ i=1,2,
\end{align*}
respectively, where
\begin{align*}
\begin{split}
\lambda_{u}^{1} &= \theta_{u}\left( c_{u}-d_{u}\right) +d_{u} \\
\lambda_{u}^{2} &= a_{u}+c_{u} \\
\theta_{u} &= \frac{\mathrm{e}^{-\int _{0}^{u}\left( a_{v}+c_{v}\right)\d v}}{\frac{c_{u}-d_{u}}{a_{u}+c_{u}-d_{u}} \mathrm{e}^{-\int _{0}^{u}\left( a_{v}+c_{v}\right)\d v} + \frac{a_{u}}{a_{u}+c_{u}-d_{u}} \mathrm{e}^{-\int _{0}^{u} d_{v}\d v}} ,
\end{split}
\end{align*}
with piecewise constant $a_{u}, c_{u},d_{u}> 0$, $c_{u}\neq d_{u}$. Note that if $c_{u}=d_{u}$, the generator in \eqref{eq:special} is a strong Markov structure and coincides with the $\Lambda_{u}^{\mathcal{S}}$ with $g_{u}=\min\left( \lambda_{u}^{1}, \lambda_{u}^{2}\right)$.

It can be verified that a bivariate Markov chain $X^{\mathcal{D}}=\left( X^{\mathcal{D},1},X^{\mathcal{D},2}\right)$ generated by $\left( \Lambda_{u}^{\mathcal{D}},u\geq 0\right)$,
\begin{align}
\Lambda_{u}^{\mathcal{D}} &= \begingroup
\renewcommand*{\arraystretch}{1.2}%
\renewcommand{\kbldelim}{(}
\renewcommand{\kbrdelim}{)}
\kbordermatrix{
           & (0,0)    & (0,1)   & (1,0)   &(1,1)\cr
  (0,0)  & -(a_{u}+c_{u}) & a_{u} & 0 & c_{u} \cr
  (0,1)  & 0 & -d_{u} & 0 & d_{u} \cr
  (1,0)  & 0 & 0 & -(a_{u}+c_{u}) & a_{u}+c_{u} \cr
  (1,1)  & 0 & 0 & 0 & 0 \cr
},\endgroup \label{eq:special}
\end{align}
is a weak Markov structure for $\left\lbrace Y^{1},Y^{2}\right\rbrace$. Specifically, $X^{\mathcal{D},1}$ is Markovian in its own filtration, but not the filtration of $X^{\mathcal{D}}$. Whereas $X^{\mathcal{D},2}$ is Markovian in the filtration $\mathbb{F}^{X}$, and necessarily Markovian in its own filtration.

In view of \eqref{eq:special}, we know that
\begin{equation*}
\mathbb{P}\left( X_{s}^{\mathcal{D}} = (1,0) \mid X_{t}^{\mathcal{D}} = (0,0)\right) = 0,\quad 0\leq t\leq s ,
\end{equation*}
and
\begin{equation*}
0< \mathbb{P}\left( X_{s}^{\mathcal{D},1} = 1 \mid X_{t}^{\mathcal{D}} = (0,0)\right) \neq \mathbb{P}\left( X_{s}^{\mathcal{D},1} = 1 \mid X_{t}^{\mathcal{D}} = (0,1)\right) ,\quad 0\leq t\leq s .
\end{equation*}
It means that $X^{\mathcal{D},1}$ cannot be in default alone. Saying differently, if $X^{\mathcal{D},1}$ defaults, $X^{\mathcal{D},2}$ must have defaulted at least earlier than $X^{\mathcal{D},1}$. Moreover, it holds that
\begin{equation*}
\mathbb{P}\left( X_{s}^{\mathcal{D},2} = 1 \mid X_{t}^{\mathcal{D}} = (0,0)\right) = \mathbb{P}\left( X_{s}^{\mathcal{D},2} = 1 \mid X_{t}^{\mathcal{D}} = (1,0)\right) ,\quad 0\leq t\leq s .
\end{equation*}
Whether $X^{\mathcal{D},1}$ defaults or not, it does not impact on the default of $X^{\mathcal{D},2}$.
The relationship between $X^{\mathcal{D},1}$ and $X^{\mathcal{D},2}$ is similar to \textit{parasitism}, and $X^{\mathcal{D},2}$ is the host. During a crisis, we would like to prevent $X^{\mathcal{D},2}$ from default to avoid Armageddon. Financially, we say that $X^{\mathcal{D},2}$ is systemically more important than $X^{\mathcal{D},1}$. One remark should be made that this relationship between $X^{\mathcal{D},1}$ and $X^{\mathcal{D},2}$ is embedded in the structure of $\Lambda_{u}^{\mathcal{D}}$ given in \eqref{eq:special}. It is independent of functions $a,c,d$.

\begin{table}[h]
\centering
\caption{Parameters for Example \ref{ex:scmSysImp} (Systemic Importance)}\label{tbl:parEx4}
\renewcommand{\arraystretch}{1.2}
\begin{tabular}{{l}*{7}{c}c}
\hline \hline
\multirow{2}{*}{Parameters} & \multicolumn{7}{c}{Time Periods} \\
\cline{3-9}
& & $[0,6)$ & $[6,10)$  & $[10,20)$ & $[20,26)$ & $[26,30)$ & $[30,\infty)$  &  \\
\hline
\multirow{3}{*}{Scenario 1:} & $a_{u}$ & 0.02 & 0.02 & 0.02 & 0.02 & 0.02 & 0.02   \\
               & $c_{u}$ & 0.02 & 0.09 & 0.06 & 0.02 & 0.09 & 0.09 &  \\
               & $d_{u}$ & 0.01 & 0.01 & 0.01 & 0.01 & 0.01 & 0.01 &    \\
\hline
\multirow{3}{*}{Scenario 2:} & $a_{u}$ & 0.02 & 0.02 & 0.02 & 0.02 & 0.02 & 0.02   \\
               & $c_{u}$ & 0.09 & 0.06 & 0.02 & 0.09 & 0.02 & 0.02 &   \\
               & $d_{u}$ & 0.01 & 0.01 & 0.01 & 0.01 & 0.01 & 0.01 &    \\
\hline
\end{tabular}
\end{table}

Now, we turn our attention to the numerical results. In what follows, we would like to study the effect of the \textit{second distress}. The parameters are provided in Table \ref{tbl:parEx4}. Both scenarios have the same parameters in $a_{u}$ and $d_{u}$, but different $c_{u}$ cycling. In scenario 1, the simulated system embarks from good condition, in the sense of the magnitude of $c_{u}$, while in scenario 2 the simulated system starts from bad condition. In Figure \ref{fig:scmEx4-1} on page \pageref{fig:scmEx4-1} and Figure \ref{fig:scmEx4-2} on page \pageref{fig:scmEx4-2}, $X^{\mathcal{D}}$ produces the highest level in systemic instability in both scenarios. This unsymmetrical relationship between two components is considered more risky than the other dependence structures, which allow individual default symmetrically. Next observation is regarding the second distress occurred in period $[26,30)$ of scenario 1 and in period $[20,26)$ of scenario 2. Although the impact of the second distress is inconclusive\footnote{We may have higher systemic instability in the second distress with different time periods setting. For instance, see Figure \ref{fig:scmEx4-3} on page \pageref{fig:scmEx4-3} by using the same parameters of scenario 1 in Table \ref{tbl:parEx4} with time periods $[0,6)$, $[6,8)$, $[8,10)$, $[10,12)$, $[12,15)$, $[15,\infty)$.}, the systemic instability in scenario 2 is indeed higher if the system starts from the bad condition.

\begin{figure}[!htb]
\centering
\includegraphics[width=5in]{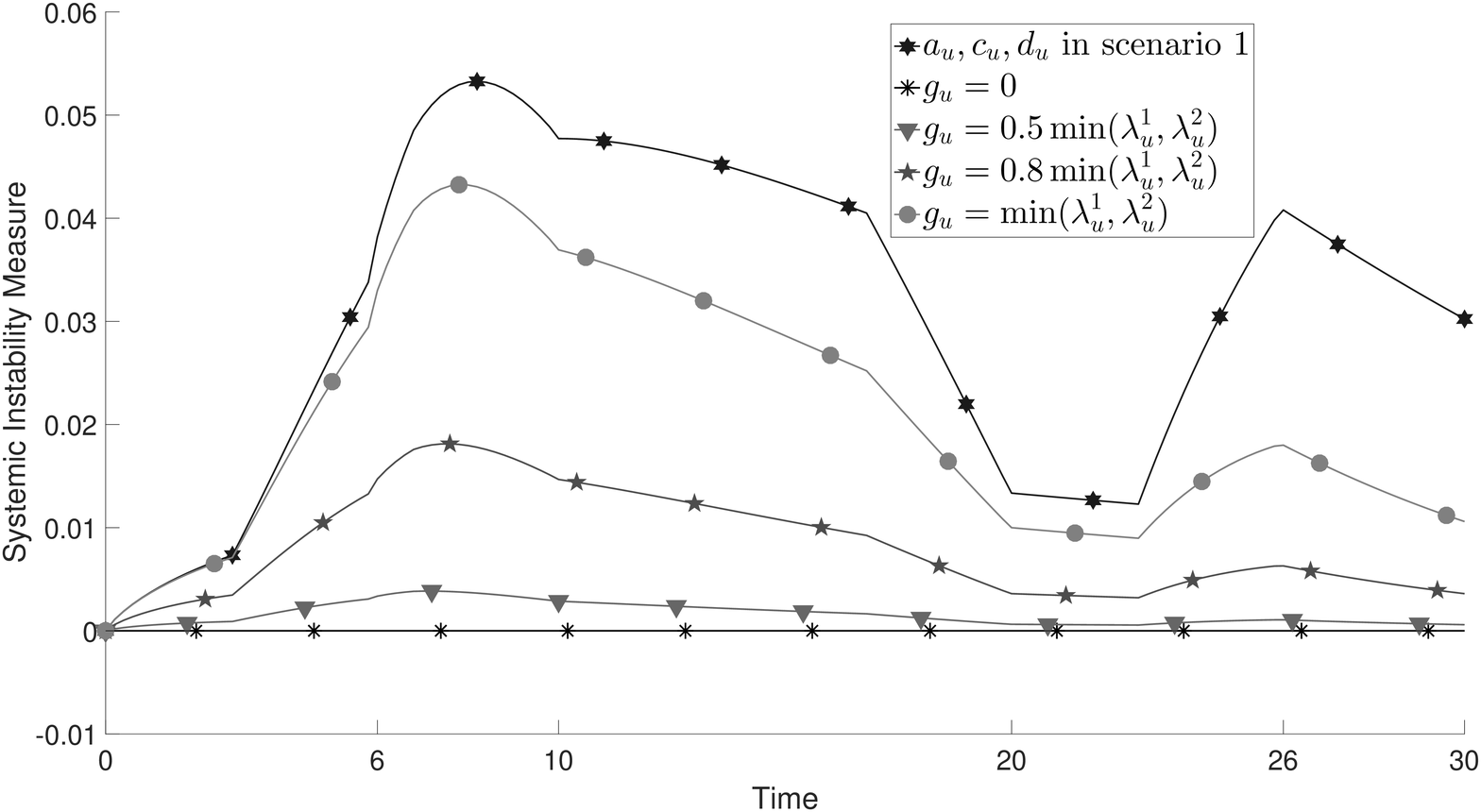}
\caption{Systemic Instability Measure for Example \ref{ex:scmSysImp} (Systemic Importance): Scenario 1}\label{fig:scmEx4-1}
\end{figure}

\begin{figure}[!htb]
\centering
\includegraphics[width=5in]{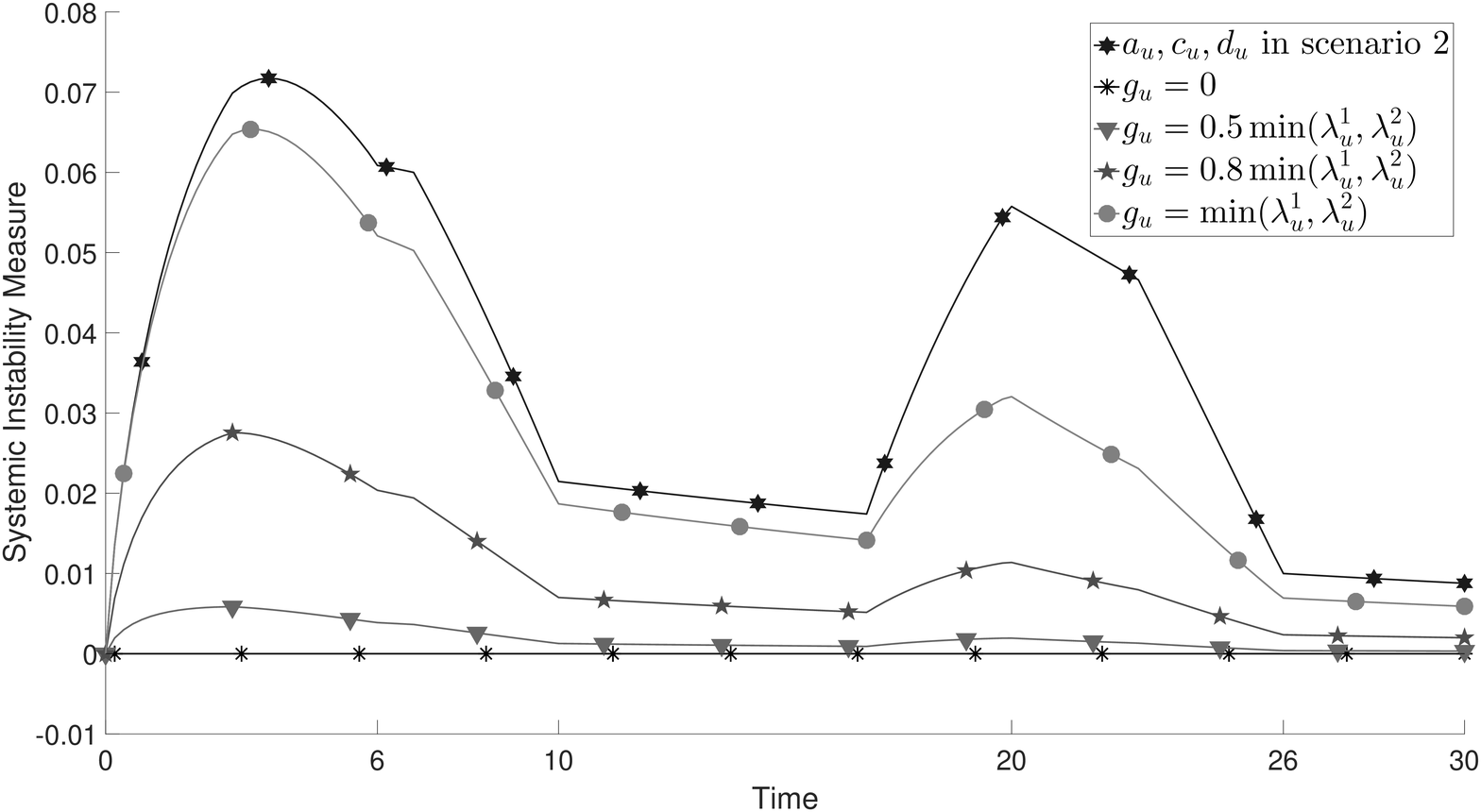}
\caption{Systemic Instability Measure for Example \ref{ex:scmSysImp} (Systemic Importance): Scenario 2}\label{fig:scmEx4-2}
\end{figure}

\begin{figure}[!htb]
\centering
\includegraphics[width=5in]{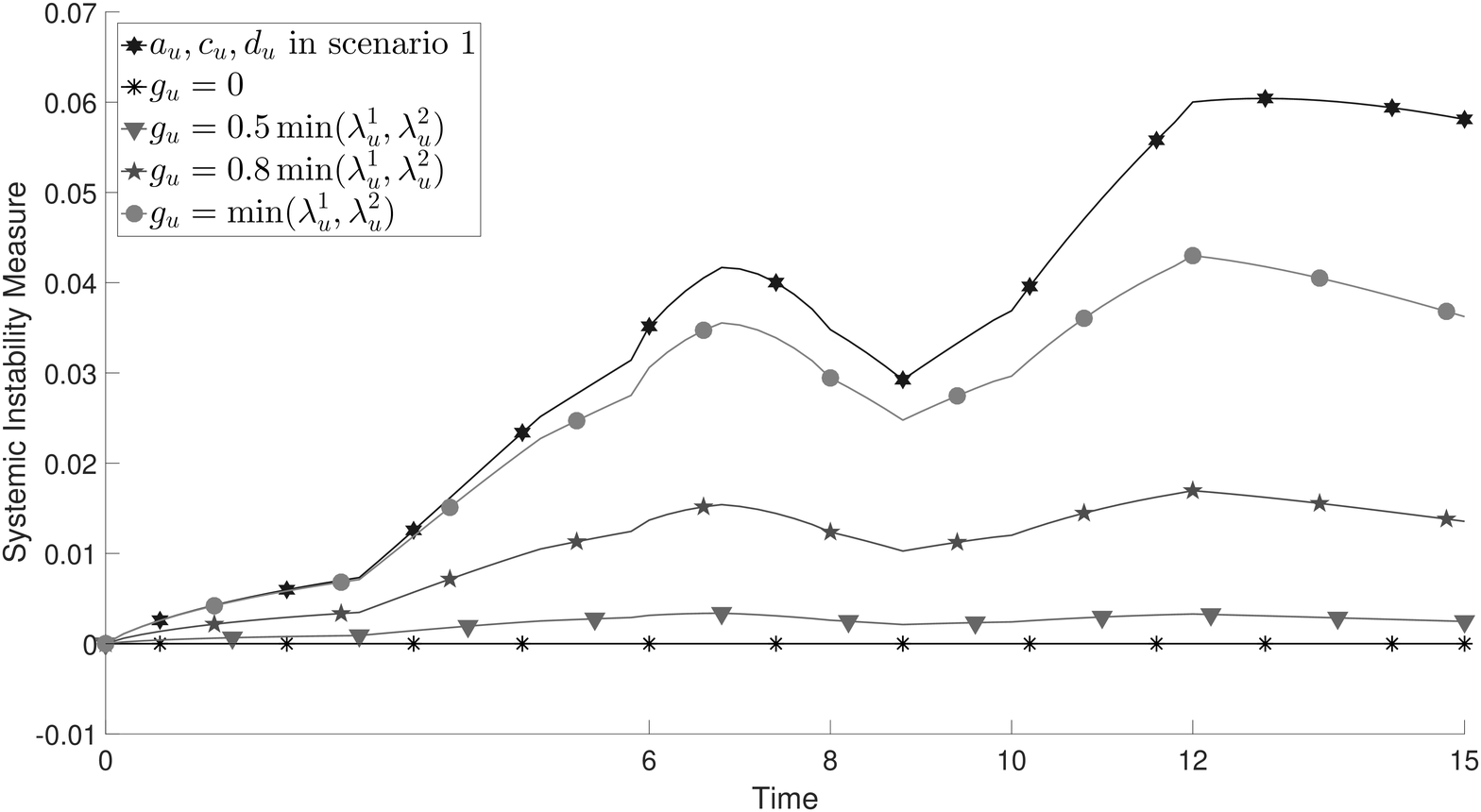}
\caption{Systemic Instability Measure for Example \ref{ex:scmSysImp} (Systemic Importance): Scenario 1 with Time Periods $[0,6),[6,8),[8,10),[10,12),[12,15),[15,\infty)$.}\label{fig:scmEx4-3}
\end{figure}

\end{example}

\begin{example}[Two weak-only Markov strcutures]\label{ex:scmCompare}
In this final example, we will compare systemic instability measures of two weak-only Markov structures for the same collection of identical Markov chains $\lbrace Y^{1},Y^{2}\rbrace$. Specifically, we take $a=b$ in \eqref{eq:srmEx1-1} and obtain $\lambda_{u}^{1} = \lambda_{u}^{2}$. We use this simple example to show that a financial system with distinct dependence structures produces different levels of systemic instability. In addition, the results are consistent with financial intuitions.

The first weak-only Markov structure $\left( \Lambda_{u}^{\mathcal{D}}, u\geq 0\right)$ is of the form
\begin{align}
\Lambda_{u}^{\mathcal{D}} &= \begingroup
\renewcommand*{\arraystretch}{1.2}%
\renewcommand{\kbldelim}{(}
\renewcommand{\kbrdelim}{)}
\kbordermatrix{
           & (0,0)    & (0,1)   & (1,0)   &(1,1)\cr
  (0,0)  & -(2a+c_{u}) & a & a & c_{u} \cr
  (0,1)  & 0 & -a & 0 & a \cr
  (1,0)  & 0 & 0 & -a & a \cr
  (1,1)  & 0 & 0 & 0 &  0\cr
},\endgroup \label{eq:srmEx5-com}
\end{align}
and the second weak-only Markov structure comes from Example \ref{ex:scmExtrCont} with $\left( \Lambda_{u}^{\mathcal{D}}, u\geq 0\right)$ of the form
\begin{align}
\Lambda_{u}^{\mathcal{D}} &= \begingroup
\renewcommand*{\arraystretch}{1.2}%
\renewcommand{\kbldelim}{(}
\renewcommand{\kbrdelim}{)}
\kbordermatrix{
           & (0,0)    & (0,1)   & (1,0)   &(1,1)\cr
  (0,0)  & -\lambda_{u}^{1} & 0 & 0 & \lambda_{u}^{1} \cr
  (0,1)  & 0 & 0 & 0 & 0 \cr
  (1,0)  & 0 & 0 & 0 & 0 \cr
  (1,1)  & 0 & 0 & 0 & 0 \cr
}.\endgroup \label{eq:srmEx5-extr}
\end{align}
Both Markov structures \eqref{eq:srmEx5-com} and \eqref{eq:srmEx5-extr} are the solutions of 
\begin{equation*}
\Lambda_{u}^{i} = \Theta_{u}^{i} \Lambda_{u}^{\mathcal{D}} \Phi^{i},\quad u\geq 0,\ i=1,2,
\end{equation*}
with $\Lambda_{u}^{1}=\Lambda_{u}^{2}$.

The parameters are listed in Table \ref{tbl:parEx5} on page \pageref{tbl:parEx5}. In Figure \ref{fig:scmEx5} on page \pageref{fig:scmEx5}, we observe that the $X^{\mathcal{D}}$ with extreme contagious structure produces higher values in systemic instability measure in both scenarios. This result is consistent with financial intuitions. If two financial institutions only default simultaneously, as we are measuring the event of Armageddon, this financial system is certainly more unstable than allowing any financial institution to default individually. Because the financial system allowing single default will be relatively away from Armageddon. Moreover, the system beginning with bad condition, i.e. scenario 2, generally behaves more unstable. 

\begin{table}[h]
\centering
\caption{Parameters for Example \ref{ex:scmCompare} (Two Weak-only Markov Structures)}\label{tbl:parEx5}
\renewcommand{\arraystretch}{1.2}
\begin{tabular}{{l}*{7}{c}c}
\hline \hline
\multirow{2}{*}{Parameters} & \multicolumn{7}{c}{Time Periods} \\
\cline{2-8}
& $[0,6)$ & $[6,10)$  & $[10,20)$ & $[20,26)$ & $[26,30)$ & $[30,\infty)$  &  \\
\hline
$a=b$ & 0.01 & 0.01 & 0.01 & 0.01 & 0.01 & 0.01 &   \\
Scenario 1: $c_{u}$ & 0.02 & 0.09 & 0.06 & 0.02 & 0.09 & 0.09 &  \\
Scenario 2: $c_{u}$ & 0.09 & 0.06 & 0.02 & 0.09 & 0.02 & 0.02 &   \\
\hline
\end{tabular}
\end{table}

\begin{figure}[!htb]
\begin{subfigure}[h]{0.49\linewidth}
\includegraphics[width=3.5in]{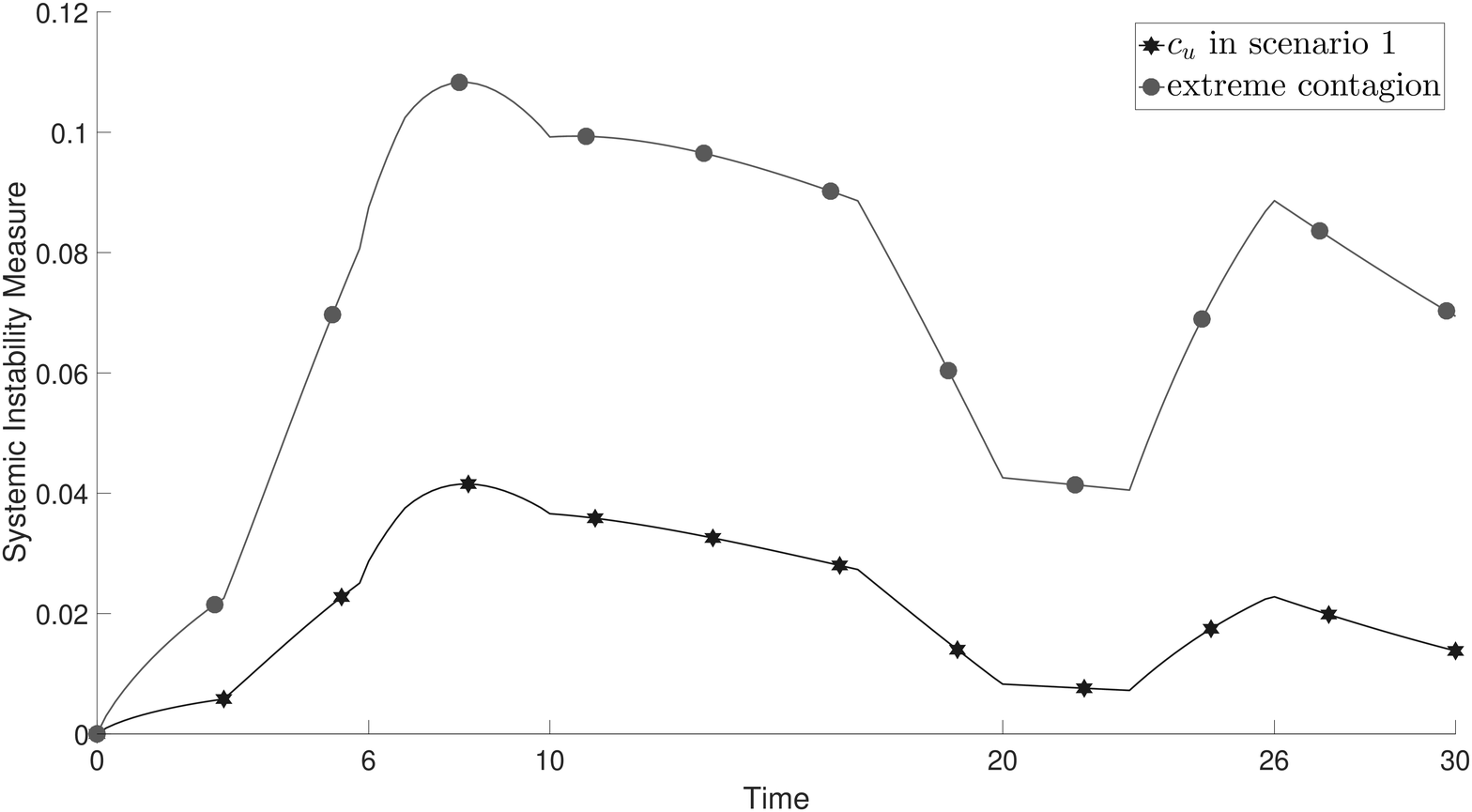}
\caption{Scenario 1}
\end{subfigure}
\hfill
\begin{subfigure}[h]{0.49\linewidth}
\includegraphics[width=3.5in]{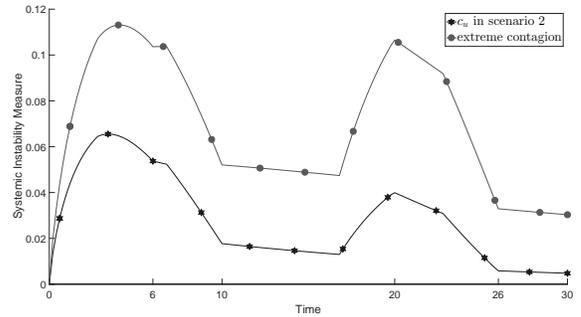}
\caption{Scenario 2}
\end{subfigure}
\centering
\caption{Systemic Instability Measure for Example \ref{ex:scmCompare} (Two Weak-only Markov Structures)}\label{fig:scmEx5}
\end{figure}

\end{example}

\begin{appendices}
\clearpage
\section{Technical Results}\label{apd:results}


In this section, we borrow condition \ref{cnd:m} from \cite{bjn13intricacies} and collect relevant results from \cite{ysc17phd} to lay the basis for our work. Recall the setting in Section \ref{sec:preliminary}, let $X=\left( X^{1},\ldots ,X^{m}\right)$ be a multivariate Markov chain and take values in $E=E_{1}\times\cdots \times E_{m}$. First, we state several conditions.

\noindent
\textbf{Condition (M)}\labeltext{(M)}{cnd:m}. \textit{We say the generator matrix function $\Lambda _{t}$ satisfies condition (M) if $\Lambda_{t}$ satisfies for every $t\geq 0$, $i=1,2,\ldots ,m$,\labeltext{(M$i$)}{cnd:mi}
\begin{equation*}\label{eq:suffCondStrong}
\left( \mathbf{M}\mathit{i}\right)\quad \sum _{y\in\mathcal{H}\left( y^{i}\right)} \lambda_{t}^{x,y} = \sum _{y\in\mathcal{H}\left( y^{i}\right)} \lambda_{t}^{\hat{x},y},\quad x^{j},\hat{x}^{j},y^{j}\in E_{j},\ j=1,2,\ldots ,m,\ x,\hat{x}\in\mathcal{H}\left( x^{i}\right),\ x^{i}\neq y^{i} .
\end{equation*}
Specifically,
\begin{equation*}
x=\left( x^{1},\ldots ,x^{i},\ldots ,x^{m}\right),\ \hat{x}=\left( \hat{x}^{1},\ldots ,x^{i}, \ldots ,\hat{x}^{m}\right),\ y=\left( y^{1},\ldots ,y^{i}, \ldots ,y^{m}\right).
\end{equation*}}

\noindent
\textbf{Condition (P)}\labeltext{(P)}{cnd:p}. \textit{We say the transition probability function $\mathbf{P}_{t,s}$ satisfies condition (P) if for any $0\leq t\leq s<\infty$, $i=1,2,\ldots ,m$,\labeltext{(P$i$)}{cnd:pi} we have
\begin{equation*}\label{def:condP}
\left( \mathbf{P}\mathit{i}\right)\quad \sum_{y\in\mathcal{H}\left( y^{i}\right)} \mathbf{P}_{t,s}^{x,y} = \sum_{y\in\mathcal{H}\left( y^{i}\right)}  \mathbf{P}_{t,s}^{\hat{x},y} ,\quad x^{j},\hat{x}^{j},y^{j}\in E_{j},\ j=1,2,\ldots ,m,\ x,\hat{x}\in\mathcal{H}\left( x^{i}\right),\ x^{i}\neq y^{i} .
\end{equation*}}
It was proved in \cite[Proposition A.2]{ysc17phd} that condition \ref{cnd:p} is equivalent to condition \ref{cnd:m}.

\begin{theorem}\label{thm:neceCondM}
Assume that
\begin{equation}\label{eq:SMCimplyM}
\mathbb{P}\left( X_{t}=z\right)>0,\ dt\operatorname{-a.e.} ,\quad z\in E.
\end{equation}
Then strong Markovian consistency implies condition \ref{cnd:m}.
\end{theorem}

We denote by $\Pi_{t}$ the collection of all finite partitions of interval $[0,t]$, namely,
\begin{equation*}
\Pi_t = \left\lbrace \pi_{t} =\lbrace 0,t_{1},\ldots ,t_{n-1},t\rbrace \mid 0=t_{0}< t_{1}< \cdots < t_{n}=t,\ n\in\mathbb{N}\right\rbrace .
\end{equation*}
We denote by $E_{i}^{n}$ the $n$-fold Cartesian product of $E_{i}$. Let $\psi(a)$ be a vector in $E_{i}^{n+1}$ with the last element in $a$. Moreover, $\Upsilon^{i}$ denotes the projection operator acting on the initial distribution,
\begin{equation*}
\left(\Upsilon^{i} \mu^{X}\right) \left( \Gamma\right) =\mathbb{P}\left( X_{0}^{i}\in\Gamma^{i}\right) ,\quad \Gamma \subset E,\ \Gamma^{i} \subset E_{i} .
\end{equation*}

\begin{theorem}\label{thm:equiMarkov}
The component $X^{i}$ of $X$ is a Markov chain in its natural filtration if and only if for $0\leq t\leq s$, any $x_{t}^{i},x_{s}^{i}\in E_{i}$, $\psi(x_{t}^{i})\in E_{i}^{n+1}$, any $\pi_{t}\in\Pi_{t}$, the following holds,
\begin{equation}\label{eq:MarkovIdentity}
\sum_{x_{t}\in\mathcal{H}\left( x_{t}^{i}\right)} \Xi^{i} \left( t,x_{t},\psi(x_{t}^{i}),\pi _{t} ,\mu^{X}\right) \mathbf{P}_{t,s}^{x_{t}, x_{s}^{i}} =0,
\end{equation}
where
\begin{align*}
\Xi^{i} \left( t,x_{t},\psi(x_{t}^{i}), \pi _{t} ,\mu^{X}\right) &:= \mathbb{P}\left( X_{t} = x_{t} \mid X_{t}^{i} = x_{t}^{i},\ X_{t_{n-1}}^{i} =x_{n-1}^{i},\ldots ,\ X_{0}^{i}=x_{0}^{i}\right) -\mathbb{P}\left( X_{t} = x_{t} \mid X_{t}^{i} = x_{t}^{i}\right) \\
\mathbf{P}_{t,s}^{x_{t},x_{s}^{i}} &:= \mathbb{P} \left( X_{s}^{i}= x_{s}^{i} \mid X_{t}= x_{t} \right) .
\end{align*}
\end{theorem}

We state the condition \ref{cnd:ci2}.

\textbf{Condition (C$i$)}\labeltext{(C$i$)}{cnd:ci2}. Given $\mu^{X}>0$, for any $t\geq 0$, $x_{t}^{i}\in E_{i},\ x_{t}\in\mathcal{H}\left( x_{t}^{i}\right),\ \psi(x_{t}^{i})\in E_{i}^{n+1}$, and $\pi_{t}\in\Pi_{t}$,
\begin{equation*}
\Xi^{i} \left( t,x_{t},\psi(x_{t}^{i}),\pi _{t} ,\mu^{X}\right) =0.
\end{equation*}

\begin{theorem}\label{thm:semiRP}
Assume that a multivariate Markov chain $X=\left( X^{1},\ldots ,X^{m}\right)$ has infinitesimal generator $\left( \Lambda_{u}, u\geq 0\right)$ with transition semigroup $\left( \mathbf{P}_{t,s}, 0\leq t\leq s\right)$. If there exists an initial distribution $\mu^{X}>0$ such that the transition semigroup $\widehat{\mathbf{P}}_{t,s}^{i}$ of $\left( \Theta _{u}^{i}\Lambda _{u} \Phi ^{i}, u\geq 0\right)$ satisfies the identity
\begin{equation}\label{eq:assumption}
\Theta_{t}^{i} \mathbf{P}_{t,s} = \widehat{\mathbf{P}}_{t,s}^{i} \Theta_{s}^{i} ,\quad 0\leq t\leq s,
\end{equation}
then 
\begin{enumerate}[label=(\roman*)]
\item $\mathbf{P}_{t,s}^{i} = \widehat{\mathbf{P}}_{t,s}^{i}$, where $\mathbf{P}_{t,s}^{i}$ is the semigroup of $X^{i}$;
\item condition \ref{cnd:ci2} holds true.
\end{enumerate}
\end{theorem}

\begin{corollary}
If the assumptions in Theorem \ref{thm:semiRP} hold true, then the component $X^{i}$ of $X$ is Markov in its natural filtration with the initial distribution $\Upsilon^{i} \mu^{X}$. Moreover, $X^{i}$ has the generator $\left( \Theta _{t}^{i}\Lambda _{t} \Phi ^{i}, t\geq 0\right)$.
\end{corollary}

We conclude this section by presenting the conditions for $X$ to be weak-only Markovian consistent with respect to the component $X^{i}$.

\begin{theorem}\label{thm:weakOnly}
Suppose that $X$ is a multivariate Markov chain generated by $\left( \Lambda_{t},t\geq 0\right)$ with initial distribution $\mu^{X}>0$. If the following conditions are satisfied,
\begin{enumerate}[label=(\roman*)]
\item for any $0\leq t\leq s$, $x_{t}^{i},x_{s}^{i}\in E_{i}$, $\psi\left( x_{t}^{i}\right)\in E_{i}^{n+1}$, and $\pi_{t}\in\Pi_{t}$, Equation \eqref{eq:MarkovIdentity} holds true;

\item the inequality holds,
\begin{equation*}
\mathbb{P}\left( X_{t}=x\right)>0,\ dt\operatorname{-a.e.} ,\quad x\in E;
\end{equation*}

\item condition \ref{cnd:mi} (or equivalently, condition \ref{cnd:pi}) does not hold,
\end{enumerate}
then the component $X^{i}$ of $X$ is a Markov chain in its natural filtration, but not a Markov chain in the filtration of $X$. Necessarily, the initial distribution of $X^{i}$ is $\Upsilon^{i}\mu^{X}$ and the generator of $X^{i}$ coincides with $\left( \Theta_{t}^{i} \Lambda_{t} \Phi^{i}, t\geq 0\right)$.
\end{theorem}
\end{appendices}

\section*{Acknowledgments}
The author greatly thanks Prof. Tomasz R. Bielecki and Prof. Igor Cialenco for their enlightening comments and helpful discussions. 



\bibliographystyle{alphaabbr}

\nocite{wentzell1981,dynkin1965bookVol1,dynkin1965bookVol2}


\end{document}